\documentclass[aps,prd,nofootinbib,floatfix,twocolumn]{revtex4}

\usepackage{bm}
\usepackage{latexsym}
\usepackage{dcolumn}
\usepackage{amsfonts,amssymb,amsmath}
\usepackage{graphicx,epsfig}
\usepackage{psfrag}  
\usepackage{multirow}

\def\beq{\begin{equation}}
\def\eeq{\end{equation}}
\def\bseq{\begin{subequations}}
\def\eseq{\end{subequations}}
\def\br{\begin{eqnarray}}
\def\er{\end{eqnarray}}
\def\benu{\begin{enumerate}}
\def\eenu{\end{enumerate}}
\def\bit{\begin{itemize}}
\def\eit{\end{itemize}}

\def\l{\left}
\def\r{\right}
\def\nn{\nonumber} 
\def\pa{{\partial}}
\def\f{\frac}
\def\sq{\sqrt}
\def\n{{\nabla}}
\def\d{{\rm d}}
\def\MP{M_{_{\rm P}}}

\reversemarginpar

\begin{document}


\title{An introduction to inflation and cosmological perturbation 
theory\footnote{I will restrict my discussion to {\it linear}\/ 
perturbation theory.
Considering higher order perturbations, as is done, say, while 
studying non-Gaussianities, though it is currently attracting a lot 
of attention, is beyond the scope of this introductory review.}}
\author{L.~Sriramkumar}\email[E-mail: ]{sriram@hri.res.in}
\affiliation{Harish-Chandra Research Institute, Chhatnag Road,
Jhunsi, Allahabad 211 019, India.}
\date{\today}
\begin{abstract}
This article provides an introductory review of inflation and 
cosmological perturbation theory. 
I begin by motivating the need for an epoch of inflation during 
the early stages of the radiation dominated era, and describe 
how inflation is typically achieved using scalar fields.
Then, after an overview of linear cosmological perturbation theory, 
I derive the equations governing the perturbations, and outline the 
generation of the scalar and the tensor perturbations during inflation.
I illustrate that slow roll inflation naturally leads to an almost 
scale invariant spectrum of perturbations, a prediction that seems 
to be in remarkable agreement with the measurements of the 
anisotropies of the cosmic microwave background.
I describe the constraints from the recent observations on some of the 
more popular models of inflation.
I conclude with a brief discussion on the status and certain prospects 
of the inflationary paradigm.
\end{abstract}
\maketitle


\section{Introduction}


\subsection{A major drawback of the hot big bang model}

The prevailing theory about the origin and the evolution of our 
universe is the so-called hot big bang model.
The model is based on two crucial observations: the discovery of the 
expansion of the universe as characterized by the Hubble's law, and 
the existence of an exceedingly isotropic and a perfectly thermal 
Cosmic Microwave Background (CMB) radiation.
Since the energy density of radiation falls faster with the expansion 
than that of matter, these two observations immediately point to the 
fact that the universe has expanded from a hot and dense early phase, 
when radiation, rather than matter, was dominant.
The transition to the more recent matter dominated epoch occurs when 
the radiation density falls sufficiently low that the photons cease 
to interact with matter.
The CMB is nothing but the relic radiation which is reaching us today 
from this epoch of decoupling.
While fairly isotropic, the CMB possesses small anisotropies (of about 
one part in $10^{5}$), and the observed pattern of the fluctuations in
the CMB provides a direct snapshot of the universe at this epoch.
The hot big bang model has been rather successful in predicting, say, 
the primordial abundances of the light elements in terms of a single 
parameter, viz. the baryon-to-photon ratio, and the value required to 
fit these observations matches the value that has been arrived at 
independently from the structure of the anisotropies in the CMB.
Despite the success of the hot big bang model in explaining the results 
from different observations, the model has a serious drawback.
Under the model, the CMB photons arriving at us today from sufficiently 
widely separated directions in the sky could not have interacted at the 
time of decoupling.
Nevertheless, one finds that the temperature of the CMB photons reaching
us from any two diametrically opposite directions hardly differ.
(For a detailed description of the various successes and the few shortcomings 
of the hot big bang model, and the original references for the different 
points mentioned above, I would refer the reader to the following
texts~\cite{kolb-1990,linde-1990,liddle-1999,paddy-2002,dodelson-2003,mukhanov-2005,weinberg-2008,durrer-2008}.)


\subsection{The scope and success of inflation}

Inflation---which refers to a period of accelerated expansion during 
the early stages of the radiation dominated epoch---provides a 
satisfactory resolution to above-mentioned shortcoming of the hot big
bang model.
In fact, in addition to offering an elegant explanation for the extent 
of homogeneity and isotropy of the background universe, inflation also 
provides an attractive causal mechanism to generate the inhomogeneities
superimposed upon it. 
The inflationary epoch amplifies the tiny quantum fluctuations present 
at the beginning of the epoch and converts them into classical perturbations 
which leave their imprints as anisotropies in the CMB. 
These anisotropies in turn act as seeds for the formation of the large 
scale structures that we observe at the present time as galaxies and 
clusters of galaxies. 
With the anisotropies in the CMB being measured to greater and greater
precision, we have an unprecedented scope to test the predictions of 
inflation. 
As I shall discuss, the simplest models of inflation driven by a single, 
slowly rolling scalar field, generically predict a nearly scale invariant 
spectrum of primordial perturbations, which seems to be in excellent 
agreement with the recent observations of the CMB.
(For a discussion on these different aspects of the inflationary paradigm, 
in addition to the texts~\cite{kolb-1990,linde-1990,liddle-1999,paddy-2002,dodelson-2003,mukhanov-2005,weinberg-2008,durrer-2008}, 
see the following reviews~\cite{kodama-1984,brandenberger-1985,mukhanov-1992,durrer-1994,lidsey-1997,lyth-1999,riotto-2002,kinney-2003,durrer-2004,martin-2004,giovannini-2005,bassett-2006,giovannini-2007,straumann-2006,kinney-2009}.)


\subsection{The organization of this review}

This article presents an introductory survey of inflation and linear
cosmological perturbation theory, and it is organized as follows.
In the next section, I shall describe the main shortcoming of the hot
big bang model, viz. the horizon problem.
In Sec.~\ref{sec:i}, I shall outline how inflation helps to overcome the 
horizon problem, and illustrate how inflation can be driven with scalar 
fields.
I shall also introduce the concept of slow roll inflation, and discuss the 
solutions to the background equations for a certain class of inflationary 
models in the slow roll approximation.
In Sec.~\ref{sec:elcpt}, I shall present an overview of linear perturbation 
theory.
I shall explain the classification of the perturbations into scalars,
vectors and tensors, and derive the equations governing these perturbations.
I shall also discuss the behavior of the scalar perturbations in a 
particular limit.
In Sec.~\ref{sec:gpdi}, after describing the generation of perturbations 
during inflation, I shall calculate the scalar and the tensor spectra that
arise in power law as well as slow roll inflation.
In Sec.~\ref{sec:cwocim}, I shall discuss how such spectra compare with 
the recent observations, and point out the constraints on some of the 
more popular models of inflation.
Finally, in Sec.~\ref{sec:sp}, I shall conclude with a brief discussion 
on the status and some prospects of the inflationary paradigm.


\subsection{Conventions and notations}

Before I get down to brass tacks, let me list the various conventions 
and notations that I shall adopt.
Unless I mention otherwise, I shall work in $(3 + 1)$-dimensions, and
I shall adopt the metric signature of $(+,-,-,-)$.
While Greek indices shall denote all the spacetime coordinates, the Latin 
indices shall refer to the spatial coordinates.
I shall set $\hbar=c=1$, but shall display $G$ explicitly, and define the 
Planck mass to be $\MP=\l(8\, \pi\, G\r)^{-1/2}$.
I shall express the various quantities in terms of either the cosmic
time~$t$ or the conformal time $\eta$, as is convenient.
An overdot and an overprime shall denote differentiation with respect 
to the cosmic and the conformal time coordinates of the Friedmann metric 
that describes the expanding universe.
It is useful to note here that, for any given function, say, $f$,
${\dot f}=(f'/a)$ and ${\ddot f}=\l[\l(f''/a^2\r)-\l(f'\,
a'/a^{3}\r)\r]$, where $a$ is the scale factor associated with the 
Friedmann metric.
Lastly, since observations indicate that the universe has a rather 
small curvature, as it is usually done in the context of inflation, 
I shall work with the spatially flat Friedmann model.

 
\subsection{A few words on the references}

The different texts~\cite{kolb-1990,linde-1990,liddle-1999,paddy-2002,dodelson-2003,mukhanov-2005,weinberg-2008,durrer-2008} 
and the variety of
reviews~\cite{kodama-1984,brandenberger-1985,mukhanov-1992,durrer-1994,lidsey-1997,lyth-1999,riotto-2002,kinney-2003,durrer-2004,martin-2004,giovannini-2005,bassett-2006,straumann-2006,giovannini-2007,kinney-2009} that I have already mentioned discuss the motivations for the inflationary scenario, the different models of inflation, the formulation 
of cosmological perturbation theory, and also how the various models compare 
with the recent observations.
In what follows, I shall typically refer to one or more of these texts or 
reviews.
I have also tried to refer to the original papers.
However, I should stress that these references are often representative and 
not necessarily exhaustive.


\section{The horizon problem}\label{sec:hp}

Let me begin by describing the horizon problem in the hot big bang model.
Consider a spatially flat, smooth, Friedmann universe described by line 
element
\beq
\d s^2 = \d t^2-a^{2}(t)\; \d{\bf x}^2
= a^{2}(\eta)\, \l(\d\eta^{2} - \d{\bf x}^2\r),\label{eq:fle}
\eeq
where $t$ is the cosmic time, $a(t)$ is the scale factor, and $\eta=\int\, 
[\d t/a(t)]$ denotes the conformal time coordinate. 
In such a background, the horizon, viz. the size of a causally connected 
region, is defined as the physical radial distance travelled by a light 
ray from the big bang singularity at $t=0$ up to a given time~$t$.
The horizon can be expressed in terms of the scale factor $a(t)$ as 
follows~\cite{mukhanov-2005,weinberg-2008}:
\beq
h(t)=a(t)\int\limits_{0}^{t}\f{\d{\tilde t}}{a\l({\tilde t}\r)}.
\eeq
Let me now compare the linear dimension of the forward and the backward
light cones at the time of decoupling in the hot big bang model.

If one assumes that the universe was dominated by non-relativistic 
matter from the time of decoupling~$t_{\rm dec}$ until today~$t_{0}$, 
then the physical size of the region on the last scattering surface 
from which we receive the CMB is given by
\beq
\ell_{_{\rm B}}\l(t_{0},t_{\rm dec}\r)
=a_{\rm dec}\, 
\int\limits_{t_{\rm dec}}^{t_{0}}\f{\d{\tilde t}}{a\l({\tilde t}\r)} 
\simeq 3\, \l(t_{\rm dec}^{2}\, t_{0}\r)^{1/3},
\eeq
where $a_{\rm dec}$ denotes the value of the scale factor at decoupling,
and I have used the fact that $t_{0}\gg t_{\rm dec}$ in arriving at the 
final expression.
On the other hand, if I assume that the universe was radiation 
dominated from the big bang until the epoch of decoupling, then 
the linear dimension of the horizon at decoupling turns out to be
\beq
\ell_{_{\rm F}}\l(t_{\rm dec},0\r)
=a_{\rm dec}\, \int\limits_{0}^{t_{\rm dec}}\,
\f{\d {\tilde t}}{a\l({\tilde t}\r)}=\l(2\, t_{\rm dec}\r).
\eeq
The ratio, say, $R$, of the linear dimensions of the backward and the 
forward light cones at decoupling then reduces to~\cite{bassett-2006}
\beq
R\equiv \l(\frac{\ell_{_{\rm B}}}{\ell_{_{\rm F}}}\r)
=\l(\f{3}{2}\r)\, \l(\f{t_{0}}{t_{\rm dec}}\r)^{1/3} 
\simeq 70,
\eeq 
and the last equality follows from the observational fact that $t_{0}
\simeq 10^{10}\; {\rm years}$, while $t_{\rm dec} \simeq 10^{5}\; 
{\rm years}$.
In other words, the linear dimension of the backward light cone is 
about $70$ times larger than the forward light cone at decoupling.
Despite this, the CMB turns out to be extraordinarily isotropic.
This is the horizon problem. 

There is another way of stating the horizon problem.
Note that the physical wavelengths associated with the perturbations, 
say, $\lambda_{_{\rm P}}$, always grow as the scale factor, i.e. 
$\lambda_{_{\rm P}} \propto a$.
In contrast, in a power law expansion of the form $a(t)\propto t^{q}$, 
the Hubble radius\footnote{I consider the Hubble radius rather than 
the horizon size since, in any power law expansion, the Hubble radius 
is equivalent to the horizon up to a finite multiplicative constant.
Also, being a local quantity, the Hubble radius often proves to be more 
convenient to handle than the horizon.}, viz. $d_{_{\rm H}}=
H^{-1}=({\dot a}/a)^{-1}$, goes as $a^{1/q}$, so that we have 
$(\lambda_{_{\rm P}}/d_{_{\rm H}})\propto a^{[(q-1)/q]}$.
This implies that, when $q<1$---a condition which applies to both the
radiation and the matter dominated epochs---the physical wavelength 
grows {\it faster}\/ than the corresponding Hubble radius {\it as we 
go back in time}\/~\cite{kolb-1990}.
In other words, the primordial perturbations need to be correlated on 
scales much larger than the Hubble radius at sufficiently early times 
in order to result in the anisotropies that we observe in the CMB.
Consequently, within the hot big bang model, any mechanism that is 
invoked to generate the primordial fluctuations has to be intrinsically 
acausal.


\section{The inflationary scenario}\label{sec:i}  

Actually, apart from the horizon problem, there also exist a few 
other puzzles to which the hot big bang model is unable to provide 
a satisfactory solution.
These other issues include, the extent to which the universe happens 
to be spatially flat today (to about one part in $10^{2}$), and the 
unacceptable density of relics that would have been formed when high 
energy symmetries were broken in the early universe, to name just two.
Amongst all these issues, the horizon problem is arguably the most 
significant.
Moreover, it so happens that, the inflationary solution to the horizon
problem also aids in surmounting the other difficulties as 
well~\cite{starobinsky-1979-1980,kazanas-1980,sato-1981,guth-1981,linde-1982,albrecht-1982}.
For these reasons, I shall restrict my attention to the horizon 
problem.
In the following sub-sections, after illustrating how inflation helps 
in overcoming the horizon problem, I shall outline how inflation is 
typically achieved with scalar fields.


\subsection{Bringing the modes inside the Hubble radius}

As I discussed above, length scales of cosmological interest today 
(say, $1\, \lesssim \lambda_{0} \lesssim 10^{4}\; {\rm Mpc}$), enter 
the Hubble radius either during the radiation or the matter dominated 
epochs, and are outside the Hubble radius at earlier times.
If a causal mechanism is to be responsible for the origin of the 
inhomogeneities, then, clearly, these length scales should be inside 
the Hubble scales (i.e. $\lambda_{_{\rm P}} < d_{_{\rm H}}$) in the 
very early stages of the universe.
This will be possible provided we have an epoch in the early universe 
during which $\lambda_{_{\rm P}}$ decreases faster than the Hubble 
radius {\it as we go back in time},\/ i.e. if we 
have~\cite{dodelson-2003,kinney-2009}
\beq
-\f{\d}{\d t}\l(\frac{\lambda_{_{\rm P}}}{d_{_{\rm H}}}\r)<0,
\eeq
which then leads to the condition that 
\beq
{\ddot a}>0.
\eeq
In other words, the universe needs to undergo a phase of accelerated 
(inflationary) expansion during the early stages of the radiation 
dominated epoch if a physical mechanism is to account for the generation 
of the primordial fluctuations.

\begin{figure*}[!htbp]
\begin{center}
\resizebox{280pt}{210pt}{\includegraphics{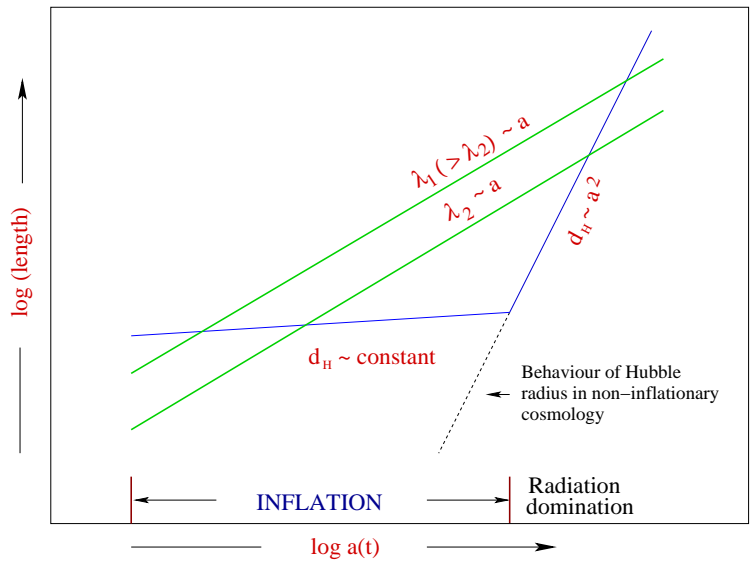}}
\caption{Evolution of the physical wavelength $\lambda_{\rm p}$ 
(in green) and the Hubble radius $d_{_{\rm H}}$ (in blue) has been 
plotted as a function of the scale factor $a$ on a logarithmic plot 
during the inflationary and the radiation dominated epochs.
As discussed in the text, in such a plot, the slope of the Hubble 
length is much less than unity during inflation, a feature which 
allows to bring around the modes inside the Hubble radius at a 
suitably early time.}
\label{fig:emdird}
\end{center}
\end{figure*}

In order to illustrate these points, in Fig.~\ref{fig:emdird}, I 
have plotted ${\rm log}\; ({\rm length})$, where `length' denotes 
either the physical wavelength of a mode or the Hubble radius, 
against ${\rm log}\; a$, during inflation and the radiation dominated 
epochs~\cite{kolb-1990}. 
For convenience, I have chosen to describe inflation by the power law 
expansion $a \propto t^{q}$, with $q>1$.
In such a situation, while $\lambda_{_{\rm P}} \propto a$ during all the 
epochs, the Hubble length $d_{_{\rm H}}$ behaves as $a^{1/q}$ and $a^{2}$ 
during inflation and the radiation dominated epochs. 
Evidently, all these quantities will be described by straight lines in 
the above plot. 
Whereas the straight lines describing the evolution of the physical 
wavelengths will have a unit slope, those describing the Hubble radius 
during the inflationary and the radiation dominated epochs will have 
a slope of much less than unity (say, for $q\gg 1$) and two, respectively.
It is then clear from the figure that, {\it as we go back in time},\/ modes 
that leave the Hubble radius during the later epochs will not be inside the 
Hubble radius during the early stages of the universe {\it unless we have 
a period of inflation.}\/


\subsection{How much inflation do we need?}

Let us now try and understand as to how much inflation we shall require
in order to ensure that the forward lightcone from the big bang to the
epoch of decoupling is at least as large as the backward light cone from
today to the epoch of decoupling, thereby resolving the horizon problem.
For simplicity, let me assume that the universe undergoes exponential 
inflation, say, from time $t_{\rm i}$ to $t_{\rm f}$, during the early 
stages of the radiation dominated epoch.
Let ${\sf H}$ be the constant Hubble scale during exponential inflation, 
and $A$ denote the extent by which the scale factor increases during the
inflationary epoch.
For $A\gg 1$, the dominant contribution to the size of the horizon at 
decoupling arises due to the rapid expansion during inflation, and it
can be evaluated to be
\beq
\ell_{_{\rm I}}\l(t_{\rm dec},0\r)
= a_{\rm dec}\, \int\limits_{0}^{t_{\rm dec}}\, 
\f{\d{\tilde t}}{a\l({\tilde t}\r)}
\simeq \l(\f{a_{\rm dec}}{\sf H}\r)\, 
\l(\f{t_{\rm dec}}{t_{\rm f}}\r)^{1/2}\, A,
\eeq
where I have set $t_{\rm i} \simeq {\sf H}^{-1}$.
In such a case, the ratio of the forward and the backward lightcones at
the epoch of decoupling is given by~\cite{paddy-2002} 
\beq
R_{_{\rm I}}
\equiv \l(\f{\ell_{_{\rm I}}}{\ell_{_{\rm B}}}\r) 
\simeq \l(\f{A}{10^{26}}\r)
\eeq
and, in arriving at the final number, I have chosen 
${\sf H}\simeq 10^{13}\; {\rm GeV}$.
Clearly, $R_{_{\rm I}}\simeq 1$, if $A\simeq 10^{26}$.
Given the scale factor, the amount of expansion that has occurred from 
an initial time $t_{\rm i}$ to a time $t$ is usually expressed in terms 
of the number of $e$-folds defined as follows: 
\beq
N=\int\limits_{t_{\rm i}}^{t} \d t\; H
={\rm ln}\; \l(\f{a(t)}{a_{\rm i}}\r).\label{eq:N}
\eeq
Since ${\rm ln}\; A ={\rm ln}\; (a_{\rm f}/a_{\rm i}) \simeq 60$, it is 
often said that one requires about $60$ $e$-folds of inflation to overcome 
the horizon problem\footnote{In fact, $60$ $e$-folds 
is roughly an observational upper bound which ensures that the 
largest scale today is inside the Hubble radius during the 
inflationary epoch~\cite{dodelson-2003b,liddle-2003}.
Actually, the number of $e$-folds needed to resolve the horizon problem 
depends on the energy scale at which inflation takes place. 
For instance, if inflation is assumed to occur at a rather low energy scale
of, say, ${\sf H}\simeq 10^{2}\, {\rm GeV}$, then, even $50$ $e$-folds will 
suffice to surmount the horizon problem.}.


\subsection{Driving inflation with scalar fields}\label{sec:diwsfs}

If $\rho$ and $p$ denote the energy density and pressure of the smooth 
component of the matter field that is driving the expansion, then the
Einstein's equations corresponding to the line element~(\ref{eq:fle}) 
result in the following two Friedmann equations for the scale factor $a(t)$:
\bseq
\label{eq:fe}
\br
H^{2}&=& \l(\f{8\,\pi\, G}{3}\r)\, {\rho},\label{eq:fe1}\\
\l(\f{\ddot a}{a}\r)
&=& -\l(\f{4\,\pi\, G}{3}\r)\, \l({\rho}+3\, p\r),\label{eq:fe2}
\er
\eseq
where, as I had mentioned, $H=\l({\dot a}/a\r)$ is the Hubble parameter.
It is clear from the second Friedmann equation that, for ${\ddot a}$ to 
be positive, we require that $\l({\rho}+3\, p\r)<0$.
Neither ordinary matter (corresponding to $p=0$) nor radiation [which 
corresponds to $p=(\rho/3)$] satisfy this condition. 
In such a situation, we need to identify another form of matter to drive 
inflation.

Scalar fields, which are often encountered in various models of high energy 
physics, can easily help us achieve the necessary condition, thereby leading 
to inflation.
Consider a canonical scalar field, say, $\phi$, described by the potential
$V(\phi)$.
Such a scalar field is governed by the action
\beq
S[\phi]=\int\! \d^{4}x\; \sqrt{-g}\, \l[\l(\f{1}{2}\r)\,
\l(\pa_{\mu}\phi\; \pa^{\mu}\phi\r)-V(\phi)\r],
\label{eq:sfa}
\eeq 
with the associated stress-energy tensor being given by
\beq
T^{\mu}_{\nu}= \pa^{\mu}\phi\; \pa_{\nu}\phi
-\delta^{\mu}_{\nu}\, \l[\l(\f{1}{2}\r)\;
\l(\pa_{\lambda}\phi\; \pa^{\lambda}\phi\r)-V(\phi)\r].
\label{eq:setsf}
\eeq
The symmetries of the Friedmann background---viz. homogeneity and 
isotropy---imply that the scalar field will depend only on time
and, hence, the resulting stress-energy tensor will be diagonal.
Therefore, the energy density~$\rho$ and the pressure~$p$ associated 
with the scalar field simplify to
\bseq
\label{eq:rhopsf}
\br
T^{0}_{0} &=&\rho=\l[\l(\f{{\dot \phi}^2}{2}\r)+V(\phi)\r],
\label{eq:rhosf}\\
T^{i}_{j} &=&-p\; \delta^{i}_{j}
=-\l[\l(\f{{\dot \phi}^2}{2}\r)-V(\phi)\r]\, \delta^{i}_{j}.
\label{eq:psf}
\er
\eseq
Moreover, from the action~(\ref{eq:sfa}), one can arrive at the 
following equation of motion for the scalar field $\phi$ in the
Friedmann universe:
\beq
{\ddot \phi}+3\, H\,{\dot \phi}+V_{\phi}=0,\label{eq:emsf}
\eeq
where $V_{\phi}=(\d V/\d\phi)$.
From the above expressions for $\rho$ and $p$, one finds that the 
condition for inflation, viz. $(\rho+3\, p)<0$, reduces to
\beq
{\dot \phi}^{2}< V(\phi).\label{eq:ci}
\eeq 
In other words, inflation can be achieved if the potential energy of the
scalar field dominates its kinetic energy.

Given a $V(\phi)$ that is motivated by a high energy model, the first 
Friedmann equation~(\ref{eq:fe1}) and the equation~(\ref{eq:emsf}) that
governs the evolution of the scalar field have to be consistently solved 
for the scale factor and the scalar field, with suitable initial conditions. 
But, using the expressions~(\ref{eq:rhopsf}) for the energy density 
and the pressure associated with the scalar field, the Friedmann 
equations~(\ref{eq:fe}) can be rewritten as
\bseq
\label{eq:HHdot}
\br
H^2 &=& \l(\frac{1}{3\, \MP^2}\r)\; 
\l[\l(\frac{{\dot \phi}^2}{2}\r)+V(\phi)\r],\label{eq:H}\\
{\dot H} &=& -\l(\frac{1}{2\, \MP^2}\r)\, {\dot \phi}^2,\label{eq:Hdot}
\er
\eseq
where, for convenience, I have set $(8\, \pi\, G)=\MP^{-2}$, as I had
defined earlier.
These two equations can then be combined to express the scalar field 
and the potential parametrically in terms of the cosmic time~$t$ as 
follows~\cite{paddy-2002}: 
\bseq
\label{eq:phiVt}
\br
\phi(t)&=&\sqrt{2}\; \MP\,
\int\! \d t\; \sqrt{\l(-{\dot H}\r)},\label{eq:phit}\\
V(t)&=& \MP^{2}\;\; \l(3\, H^{2}+{\dot H}\r).
\label{eq:Vt}
\er
\eseq
If we know the scale factor $a(t)$, these two equations allow us to 
`reverse engineer' the potential from which such a scale factor can 
arise!
Using this procedure, I shall now `reconstruct' the potentials of two
commonly considered models of inflation. 

Consider the power law expansion
\beq
a(t)=a_{1}\, t^{q}\label{eq:ple}
\eeq 
with $q>1$ corresponding to inflation, and $a_{1}$ being an arbitrary 
constant.
On substituting this scale factor in equation~(\ref{eq:phit}) and, upon 
integration, one can immediately show that the scalar field evolves as
\beq
\l(\frac{\phi(t)}{\MP}\r)
=\sqrt{(2\, q)}\;\; {\rm ln}\, 
\l[\sqrt{\l(\frac{V_{0}}{(3\, q-1)\, q}\r)}\,
\l(\frac{t}{\MP}\r)\r],\label{eq:phitple}
\eeq
where $V_{0}$ is a constant of integration.
The potential that leads to such a behavior can then be obtained 
using equation~(\ref{eq:Vt}), and the above expressions for~$a(t)$ 
and~$\phi(t)$.
It is found to be~\cite{lucchin-1985}
\beq
V(\phi)
=V_{0}\, \exp\,- \l[\sqrt{\frac{2}{q}}\, \l(\frac{\phi}{\MP}\r)\r].
\label{eq:plp}
\eeq
In a similar fashion, it is straightforward to establish that the 
potential
\beq
V(\phi)
=\l(3\,\alpha^{2}\, \beta^{2}\, \Gamma^{\kappa}\, \MP^{2}\r)\!
\l[1-\l(\frac{\kappa^2}{6}\r)\! 
\l(\frac{\MP}{\phi}\r)^{2}\r]\!
\l(\frac{\phi}{\MP}\r)^{-\kappa}\!\!,
\eeq
where $\Gamma=\sqrt{(2\,\alpha\,\kappa)}$ and $\kappa=[4\, 
(1-\beta)/\beta]$, leads to the following behavior for $a(t)$:
\beq
a(t)=a_{1}\, \exp\, \l(\alpha\, t^{\beta}\r)
\eeq
with $\alpha>0$ and $0<\beta<1$, and $a_{1}$ is again some arbitrary 
constant.
Since this scale factor grows faster than power law inflation, but 
slower than exponential expansion, it is referred to as intermediate
inflation~\cite{barrow-1990,muslimov-1990}.


\subsection{Slow roll inflation}

The condition~(\ref{eq:ci}) that the potential energy of the 
inflaton\footnote{It is common to refer to the scalar field 
that drives inflation as the inflaton.} dominates the kinetic 
energy is necessary for inflation to take place.
However, inflation is {\it guaranteed},\/ if the field 
{\it rolls slowly}\/ down the potential such that
\beq
{\dot \phi}^{2} \ll V(\phi).\label{eq:csri}
\eeq 
Moreover, it can be ensured that the field is slowly rolling for a  
sufficiently long time (to achieve the required $60$ or so $e$-folds
of inflation), provided
\beq
{\ddot \phi}\ll \l(3\, H\, {\dot \phi}\r).
\eeq
These two conditions lead to the slow roll 
approximation~\cite{steinhardt-1984,salopek-1990,liddle-1992}, which, 
as we shall see, allows one to construct analytical solutions, both for 
the background and the perturbations.
The approximation is usually described in terms of what are referred 
to as the slow roll parameters.
Two types of slow roll parameters---the potential slow roll parameters 
and the Hubble slow roll parameters---are often considered in the 
literature\footnote{Though, I should add that, nowadays, it seems to 
be more common to use the following hierarchy of Hubble flow 
functions~\cite{schwarz-2001,leach-2002}:\[\epsilon_{0} \equiv 
\l(\frac{H_*}{H }\r) 
\quad{\rm and}\quad
\epsilon_{i+1} \equiv \l(\frac{\d \ln \vert\epsilon_i\vert}{\d N}\r)\;\;
{\rm for}\;\; i\ge 0,\]with $H_{*}$ being the Hubble
parameter evaluated at some given time during inflation.}.
I shall describe these two sets of parameters below, and discuss the
solutions for the inflaton and the scale factor for a particular class
of potentials in the slow roll approximation.


\subsubsection{The potential slow roll parameters}

When provided with a potential $V(\phi)$, the potential slow roll 
approximation corresponds to requiring the following two 
dimensionless parameters~\cite{steinhardt-1984,salopek-1990,liddle-1992}:
\beq
\epsilon_{_{\rm V}} 
= \l(\!\frac{\MP^2}{2}\!\r)\, \l(\frac{V_\phi}{V}\r)^2 
\quad{\rm and}\quad
\eta_{_{\rm V}}= \MP^2\, \l(\frac{V_{\phi\phi}}{V}\r),
\label{eq:psrp}
\eeq
where $V_{\phi\phi}\equiv\l(\d^2V/\d\phi^2\r)$, to be small when 
compared to unity.
The two quantities $\epsilon_{_{\rm V}}$ and $\eta_{_{\rm V}}$ are 
referred to as the Potential Slow Roll (PSR) parameters.
It is straightforward to show that neglecting the kinetic energy term 
$({\dot \phi}^2/2)$ in the Friedmann equation~(\ref{eq:H}) and the 
acceleration term~${\ddot \phi}$ in the equation of motion~(\ref{eq:emsf}) 
for the scalar field are equivalent to the smallness of these parameters.
However, it should be emphasized that the converse is not true.
The smallness of the PSR parameters is only a necessary condition, and 
it is not sufficient to ensure that these terms can indeed be ignored.
The reason being that the PSR parameters only restrict the form of 
the potential, and not the dynamics of the solutions. 
Even if $\epsilon_{_{\rm V}}$ and $\eta_{_{\rm V}}$ are small, 
there is no assurance that inflation will take place since the 
value of $\dot{\phi}$ can be as large as possible.
Therefore, in addition to the two PSR parameters being small, the slow 
roll approximation actually requires the additional condition that the 
scalar field is moving slowly along the attractor solution determined 
by the equation: $(3\, H\, {\dot \phi})= -V_{\phi}$ (for a detailed 
discussion on this point, see Ref.~\cite{liddle-1994}).
 
In spite of this shortcoming, the PSR parameters often prove 
to be very handy.
For instance, given a potential, they immediately allow us to
determine the domains and the parameters of the potential that 
can lead to inflation.
I shall now discuss two examples to illustrate the utility of 
these parameters. 
Consider potentials of the form~\cite{linde-1983}
\beq
V(\phi)=\l(V_{0}\, \phi^{n}\r),\label{eq:lfm}
\eeq
where $V_{0}$ is a constant and $n>0$.
Let us restrict ourselves to the region $\phi>0$ wherein $V(\phi)$
is positive for all~$n$. 
It is straightforward to show that the slow roll conditions [viz. 
$(\epsilon_{_{\rm V}}, \eta_{_{\rm V}})\ll 1$] are satisfied when 
$\phi$ is much greater than $\MP$.
Since inflation occurs for such large values of the field, 
these potentials are often classified as `large field' 
models~\cite{bassett-2006}.
Now, consider the following potential which describes the 
pseudo-Nambu-Goldstone Boson
\beq
V(\phi)=\Lambda\, \l[1+{\rm cos}\,\l(\phi/f\r)\r],
\eeq
where $\Lambda$ and $f$ are constants that characterize the depth 
and the width of the potential.
This potential `naturally' leads to inflation for values of the field
that are small when compared to the Planck scale~\cite{freese-1990}.
Hence, such models are usually referred to as `small field' models.


\subsubsection{The Hubble slow roll parameters}

The Hubble Slow Roll (HSR) parameters turn out to be a better choice 
to describe the slow roll approximation than the PSR parameters since 
they do not require any additional conditions to be satisfied for the 
approximation to be valid.
The HSR parameters are called so, since they are defined in terms 
of the Hubble parameter~$H$, which is treated as a function of the 
scalar field~$\phi$~\cite{salopek-1990}.
In such a case, we can write Eq.~(\ref{eq:HHdot}) as
\beq
{\dot \phi}= -\l(2\, \MP^2\r)\, H_{\phi},\label{eq:Hphi}
\eeq
where $H_{\phi}\equiv(\d H/\d\phi)$.
This expression can then be used to rewrite the first Friedmann 
equation~(\ref{eq:H}) as follows:
\beq
H_{\phi}^2-\l(\frac{3\, H^2}{2\, \MP^2}\r)
=-\l(\frac{V}{2\, \MP^4}\r),\label{eq:hjf}
\eeq
a relation that is referred to as the Hamilton-Jacobi formulation of
inflation~\cite{salopek-1990}.

Taking $H(\phi)$ to be the primary quantity, the dimensionless HSR 
parameters $\epsilon_{_{\rm H}}$ and $\delta_{_{\rm H}}$ are defined 
as follows~\cite{liddle-1994}:
\beq
\epsilon_{_{\rm H}} = \l(2\, \MP^2\r)\, \l(\frac{H_\phi}{H}\r)^2
\quad{\rm and}\quad
\eta_{_{\rm H}} = \l(2\, \MP^2\r)\, \l(\frac{H_{\phi\phi}}{H}\r),
\eeq
where $H_{\phi\phi}\equiv\l({\d}^{2} H/\d\phi^{2}\r)$.
On using Eqs.~(\ref{eq:emsf}), (\ref{eq:Hphi}) and~(\ref{eq:hjf}), 
these two parameters can be written as 
\bseq
\label{eq:hsrp}
\br
\epsilon_{_{\rm H}}&=& \l(\frac{3\, {\dot \phi}^2}{2\, \rho}\r)
=-\l(\frac{\dot{H}}{H^2}\r),\label{eq:hsrp1}\\
\delta_{_{\rm H}}&=&- \l(\frac{\ddot \phi}{H {\dot \phi}}\r)
=\epsilon_{_{\rm H}} - \l(\frac{\dot{\epsilon}_{_{\rm H}}}{2\, 
H\, \epsilon_{_{\rm H}}}\r),\label{eq:hsrp2}
\er
\eseq
where $\rho$ is the energy density associated with the scalar field.
The following points are clear from these expressions.
Firstly, $\epsilon_{_{\rm H}} \ll 1$ is precisely the condition required 
for neglecting the kinetic energy term in the total energy of the scalar 
field.
Secondly, the limit $\delta_{_{\rm H}} \ll 1$ corresponds to the 
situation wherein the acceleration term of the scalar field can be 
ignored in Eq.~(\ref{eq:emsf}) when compared to the term 
involving the velocity.
Finally, the inflationary condition ${\ddot a}>0$ {\it exactly}\/ 
corresponds to $\epsilon_{_{\rm H}}<1$. 

It should again be emphasized that, since the smallness of the HSR 
parameters ensure that ${\dot \phi}$ is small, the HSR approximation 
implies the PSR approximation. 
But, the converse does not hold without assuming the constraint that 
the inflaton is already on the attractor~\cite{liddle-1994}.


\subsubsection{Solutions in the slow roll approximation}

Note that the equation of motion of the scalar field~(\ref{eq:emsf}) 
and the first Friedmann equation~(\ref{eq:H}) can be written in terms 
of the two HSR parameters as 
\bseq
\label{eq:Hemsfhsrp}
\br
H^2\, \l[1 - \l(\f{\epsilon_{_{\rm H}}}{3}\r)\r]
&=&  \l(\frac{V}{3\, \MP^2}\r),\\
\l(3\, H\, {\dot \phi}\r)\; 
\l[1 - \l(\f{\delta_{_{\rm H}}}{3}\r)\r] &=& -V_{\phi}.
\er
\eseq
The slow roll approximation corresponds to the situation wherein the 
HSR parameters $\epsilon_{_{\rm H}}$ and $\delta_{_{\rm H}}$ satisfy
the following conditions:
\beq
\epsilon_{_{\rm H}} \ll 1, \quad 
\delta_{_{\rm H}} \ll 1, \quad{\rm and}\quad 
\mathcal{O}\l[\epsilon_{_{\rm H}}^2,\delta_{_{\rm H}}^2,
\l(\epsilon_{_{\rm H}} \delta_{_{\rm H}}\r)\r] 
\ll \epsilon_{_{\rm H}}.\label{eq:sra}
\eeq
At the leading order in the slow roll approximation, the 
equations~(\ref{eq:Hemsfhsrp}) above reduce to
\beq
H^{2}\simeq \l(\frac{V}{3\, \MP^2}\r)\quad{\rm and}\quad
\l(3\, H\, {\dot \phi}\r)\simeq -V_{\phi}.
\label{eq:Hemsfsra}
\eeq
Being first order differential equations, given a potential, these 
equations can be easily integrated to obtain the solutions to the 
scale factor and the scalar field in the slow roll limit.
Let me now discuss the solutions in this limit to the large field 
models~(\ref{eq:lfm}) that I had considered earlier.

For the potential~(\ref{eq:lfm}), when $n\ne 4$, in the slow roll limit, 
the solution to the scalar field is given by~\cite{durrer-2008,martin-2004}
\beq
\phi^{[(4-n)/2]}(t)
\simeq \phi^{[(4-n)/2]}_{\rm i}
+\sqrt{\f{V_{0}}{3}}
\l[\f{n\, (n-4)}{2}\r]\MP\, (t-t_{\rm i}),
\eeq
whereas, when $n=4$, one finds that
\beq
\phi(t)
\simeq \phi_{\rm i}\;
\exp-\l[\sqrt{\l(V_{0}/3\r)}\; \l(4\, \MP\r)\, (t-t_{\rm i})\r]
\eeq
and, in both these solutions, $\phi_{\rm i}$ is a constant 
that denotes the value of the scalar field at some initial 
time $t_{\rm i}$.
For all $n$, the scale factor can be expressed in terms of these 
solutions for the scalar field as follows: 
\beq
a(t)
\simeq a_{\rm i}\; \exp- \l[\l(\frac{1}{2\, n\, \MP^{2}}\r)\,
\l(\phi^{2}(t)-\phi^{2}_{\rm i}\r)\r]
\eeq
with $a_{\rm i}$ being the value of the scalar factor at $t_{\rm i}$.
It is also useful to note that, in the slow roll limit, the two 
equations~(\ref{eq:Hemsfsra}) allow us to express the number of 
$e$-folds from $t_{\rm i}$ to $t$ during inflation as 
\beq
N={\rm ln}\, \l(\frac{a}{a_{\rm i}}\r)
=\int\limits_{t_{\rm i}}^{t} \d t\, H
\simeq -\l(\frac{1}{\MP^2}\r)\, 
\int\limits_{\phi_{\rm i}}^{\phi}\d\phi\, \l(\frac{V}{V_{\phi}}\r),
\eeq
where the upper limit~$\phi$ is the value of the scalar field at the 
time~$t$.
In terms of $e$-folds, for the large field models, the scalar field 
and the Hubble parameter are given by
\bseq
\br
\!\!\!\!\!\!\!\!\!\!\!\!\!\!\!\!
\phi^{2}(N)&\simeq&\l[\phi_{\rm i}^{2}
-\l(2\, \MP^2\, n\r)\; N\r],\qquad\qquad\label{eq:slfm}\\
\!\!\!\!\!\!\!\!\!\!\!\!\!\!\!\!
H^{2}(N)&\simeq & \l(\!\frac{V_{0}\, \MP^{(n-2)}}{3}\!\r)
\l[\l(\!\frac{\phi_{\rm i}}{\MP}\!\r)^{2}-\l(2\, n\, 
N\r)\r]^{(n/2)}\!\!\!\!\!\!\!\!\!.
\er
\eseq


\section{Essential linear, cosmological perturbation 
theory}\label{sec:elcpt}

Though the inflationary scenario was originally proposed to resolve 
the different puzzles related to the smooth background, it was soon 
realized that it also offers a simple mechanism to generate the 
primordial perturbations.
Before I turn to demonstrating how inflation produces these 
inhomogeneities, 
I shall provide an overview of essential cosmological perturbation 
theory.
(As I had mentioned, I will restrict myself to linear perturbations.)
In what follows, after a discussion on the classification of the 
perturbations into scalars, vectors and tensors, I shall derive 
the equations governing these perturbations.
Since the scalar perturbations are primarily responsible for the 
anisotropies in the CMB and the formation of structures, I shall 
also briefly highlight the evolution of these perturbations at 
super-Hubble scales during the radiation and the matter dominated 
epochs.


\subsection{Classification of the perturbations}

CMB observations indicate that the anisotropies at the epoch 
of decoupling are rather small (one part in $10^{5}$, as I had 
mentioned). 
If so, the amplitude of the deviations from homogeneity will be even 
smaller at earlier epochs.
This suggests that the generation and the evolution of the perturbations 
(until structures begin to form late in the matter dominated epoch),
can be studied using linear perturbation theory.

In a Friedmann background, the metric perturbations can be decomposed 
according to their behavior under local rotation of the spatial 
coordinates on hypersurfaces of constant time. 
This property leads to the classification of the perturbations as 
scalars, vectors and tensors~\cite{bardeen-1980,stewart-1990}. 
Scalar perturbations remain invariant under rotations (and, hence, can 
be said to have zero spin).
As we shall see, these are the principal perturbations that are 
responsible for the anisotropies and the inhomogeneities in the 
universe.
Vector and tensor perturbations---as their names indicate---transform 
as vectors and tensors do under rotations (and, as a result, have spins 
of unity and two, respectively).
The vector perturbations are generated by rotational velocity fields 
and, therefore, are also referred to as the vorticity modes.
Finally, the tensor perturbations describe gravitational waves, 
and it is important to note that they can exist even in the absence 
of sources~\cite{grishchuk-1974,starobinsky-1979}.


\subsubsection{The number of independent scalar, vector 
and tensor degrees of freedom}

Let us now carry out the exercise of counting the number of 
degrees of freedom associated with these different types of
perturbations~\cite{riotto-2002}.
To highlight the counting procedure, in this sub-section, I shall, 
in fact, work in arbitrary spacetime dimensions.
In $(D+1)$-dimensions, the metric tensor, being symmetric, has 
$[(D+1)\, (D+2)/2]$ degrees of freedom. 
However, not all of these degrees are independent since there 
exist $(D+1)$ degrees of freedom associated with the coordinate 
transformations.
If we eliminate these degrees, we are left with a total of $[D\, 
(D+1)/2]$ independent degrees of freedom describing the metric 
perturbations.
These degrees of freedom contain the scalar, the vector, and the 
tensor perturbations. 

Let $\delta g_{\mu\nu}$ denote the metric perturbations in the 
Friedmann universe.
The perturbed metric can be split as follows: 
\beq
\delta g_{\mu\nu}=\l(\delta g_{00}, \delta g_{0i}, \delta g_{ij}\r).
\eeq
Since it contains no running index, evidently, the perturbation 
$\delta g_{00}={\cal A}$, say, is a scalar.
As is well known, one can always decompose a vector into a gradient 
of a scalar and a vector that is divergence free.
Hence, we can write $\delta g_{0i}$ as
\beq
\delta g_{0i}=\l(\n_{i}\,{\cal B}+{\cal S}_{i}\r),
\eeq
where ${\cal B}$ is a scalar, while ${\cal S}_{i}$ is a vector that is 
divergence free, i.e. $\l(\n_{i}\,{\cal S}^{i}\r)=0$.
A similar decomposition can be carried for the quantity $\delta g_{ij}$ 
by essentially repeating the above analysis on each of the two indices.
Therefore, the quantity $\delta g_{ij}$ can be decomposed as
\br
\delta g_{ij} 
\!\!&=&\!\! \psi\; \delta_{ij}
+ \l(\n_{i}\,{\cal F}_{j} + \n_{j}\,{\cal F}_{i}\r)\nn\\
& &\qquad\,
+\l[\l(\f{1}{2}\r) \l(\n_{i}\,\n_{j}+\n_{j}\,\n_{i}\r)
-\l(\f{1}{3}\r) \delta_{ij}\, \n^{2}\r] {\cal E}\nn\\
& &\qquad\;
+\, {\cal H}_{ij},
\er
where $\psi$ and ${\cal E}$ are scalar functions, ${\cal F}_{i}$---like 
${\cal S}_{i}$ above---is a divergence free vector, and ${\cal H}_{ij}$ is 
a symmetric, traceless and transverse tensor that satisfies the conditions 
${\cal H}^{i}_{i}=0$ and $\l(\nabla_{i}\,{\cal H}^{ij}\r)=0$.
Let me again count the degrees of freedom of the perturbed metric tensor 
through the various functions I have introduced.
To describe $\delta g_{\mu\nu}$, we require the scalars ${\cal A}$, 
${\cal B}$, $\psi$ and ${\cal E}$, amounting to four degrees of freedom.
We also need the two divergence free spatial vectors ${\cal S}_{i}$ and 
${\cal F}_{i}$ that add up to $[2\, (D-1)]$ degrees.
Moreover, after we impose the traceless and transverse conditions, the 
tensor ${\cal H}_{ij}$ has 
\beq
\l[\f{D\, (D+1)}{2}\r]-(D+1)= \l[\f{(D+1)\,(D-2)}{2}\r]
\eeq 
degrees of freedom.
Upon adding all these scalar, vector and tensor degrees, I obtain 
\beq
4+2\, (D-1)+ \l[\f{(D+1)\,(D-2)}{2}\r]\!=\!\l[\f{(D+1)\,(D+2)}{2}\r]
\eeq
which, as I discussed above, are the total number of degrees of freedom
associated with the perturbed metric in $(D+1)$-dimensions.

In the same fashion, let me decompose the degrees of freedom associated 
with the coordinate transformations.
In a Friedmann background, the $(D+1)$ coordinate transformations that
relate different coordinate systems that describe the perturbed metric
can be expressed in terms of scalars, say, $\delta t$ and $\delta x$, 
as follows:
\beq
t\to \l(t+\delta t\r)
\quad {\rm and}\quad
x^{i}\to \l[x^{i}+\n^{i}\,(\delta x)\r].\label{eq:sgt}
\eeq
Similarly, one can construct coordinate transformations in terms of a
divergence free vector, say, $\delta x^{i}$, as
\beq
t\to t 
\quad {\rm and}\quad 
x^{i}\to \l(x^{i}+\delta x^{i}\r).
\label{eq:vgt}
\eeq
(It should be emphasized that these coordinate transformations are of a 
particular form, and are not completely arbitrary.
The form of these transformations is dictated by the fact that the difference 
between the coordinates are determined by the amplitude of the perturbations.
In other words, it is a gauge transformation.
I shall somewhat elaborate on this point below.)
There is no further coordinate degrees of freedom associated with the
tensor perturbations.
The two scalar quantities $\delta t$ and $\delta x$, and the divergence
free vector $\delta x^{i}$, constitute the $[2+(D-1)]=(D+1)$ degrees of 
freedom associated with the coordinate transformations.

Let me now subtract the coordinate degrees of freedom from the total 
degrees to arrive at the independent number of scalar, vector and 
tensor degrees of freedom.
Four scalar functions, viz. $({\cal A},{\cal B},\psi,{\cal E})$, were 
required to describe the perturbed metric tensor $\delta g_{\mu\nu}$.
But, there also exist two scalar degrees of freedom associated with one's
choice of coordinates, i.e. $\delta t$ and $\delta x$.
Hence, the actual number of independent scalar degrees of freedom is 
$(4-2)=2$. 
And, it is useful to notice that this is true in arbitrary spacetime 
dimensions.
I had needed two divergence free vector functions, ${\cal S}_{i}$ 
and ${\cal F}_{i}$, amounting to a total of $[2\, (D-1)]$ degrees 
of freedom, to describe the perturbed metric tensor.
If we subtract the $(D-1)$ vector degrees of freedom corresponding 
to the coordinate transformations that can be achieved through $\delta 
x^{i}$ from this number, one is left with $[2\, (D-1)]-(D-1)=(D-1)$ 
true vector degrees of freedom.
As I have already pointed out, the tensor perturbations contain $[(D+1)\,
(D-2)/2]$ independent degrees of freedom. 
Upon adding these, I obtain that
\beq
2+(D-1)+ \l[\f{(D+1)\,(D-2)}{2}\r]=\l[\f{D\,(D+1)}{2}\r]
\eeq
which is the total number of {\it independent}\/ degrees of freedom 
describing the perturbed metric tensor.

Note that, in the $(3+1)$-dimensional case of our interest, there exists 
two each of the scalar, the vector and the tensor degrees of freedom.


\subsubsection{The decomposition theorem}

I have focused above on decomposing the perturbed metric tensor 
into the different types of perturbations.
Let me now discuss the equivalent classification of the source of 
the metric perturbations, viz. the stress-energy tensor.
Since the stress-energy tensor is a symmetric two tensor just as 
the metric tensor is, the perturbed stress-energy tensor can also 
be classified as the scalar, the vector and the tensor components.
For instance, while the perturbed inflaton and perfect fluids---such 
as radiation and matter---are scalar sources, velocity fields with 
vorticity are, evidently, vector sources.
Anisotropic stresses, when the possible scalar and vector contributions 
have been eliminated, constitute a tensor source.

Given the perturbed metric tensor $\delta g_{\mu\nu}$, the corresponding 
Einstein tensor, say, $\delta G_{\mu\nu}$, can immediately be calculated 
at the same order in the perturbations.
The Einstein equations that relate the perturbed Einstein tensor to the
perturbed stress-energy tensor, say, $\delta T_{\mu\nu}$, will then lead 
to the equations governing the perturbations.
According to the decomposition theorem~\cite{weinberg-2008,kodama-1984}, 
which I shall use without proof, at the linear order in the perturbations, 
the scalar, the vector and the tensor perturbations decouple and, hence, 
they can be analyzed separately. 
In other words, each type of metric perturbation is affected by only the
same type of source.
Therefore, the three types of perturbations can be studied independently 
of each other. 


\subsubsection{Gauges}

The Friedmann line element~(\ref{eq:fle}) describes the expanding 
universe in the frame of a comoving observer.
The comoving observer is special since, it is with respect to such an 
observer that the universe appears homogeneous and isotropic.
However, there is no such uniquely preferred frame of reference in the
presence of perturbations.
A variety of coordinate choices are possible, with the only requirement
being that the metric and the coordinates reduce to the standard 
Friedmann line element in the limit when the perturbations vanish.
As a result, the difference between the various coordinates have the 
same amplitude as the perturbations themselves.
A particular choice of coordinates is called a gauge.
If one changes the coordinate system, one would obtain another metric, 
i.e. a different gauge.
The transformation from one gauge to another [such as Eqs.~(\ref{eq:sgt}) 
and~(\ref{eq:vgt})] is referred to as a gauge transformation.
There exists two approaches to studying the evolution of the perturbations.
Either one constructs gauge invariant quantities, or chooses a particular
gauge and work in the specific gauge throughout.
Being simpler and more wieldy, I shall adopt the latter approach.


\subsection{Scalar perturbations}

I shall now turn to deriving the equations of motion governing the 
three different types of perturbations.
Let me first consider the case of the scalar perturbations.

\subsubsection{The longitudinal gauge and the perturbed Einstein tensor}

As I mentioned above, I shall work in a particular gauge to describe the 
various perturbations.
A convenient gauge to describe the scalar perturbations is the longitudinal 
gauge.
If we take into account the scalar perturbations to the background 
metric~(\ref{eq:fle}), then, in this gauge, the Friedmann line element
is given by 
\beq
{\rm d}s^2 
= \l(1+2\, \Phi\r)\, \d t^2
-a^{2}(t)\; \l(1-2\, \Psi\r)\, \d{\bf x}^2,\label{eq:flesp}
\eeq
where $\Phi$ and $\Psi$ are the two independent functions that describe 
the perturbations.
(Note that this choice of gauge corresponds to ${\cal A}\propto \Phi $,  
$\psi\propto \Psi$ and ${\cal B}={\cal E}=0$.)
At the linear order in the perturbations, the components of the 
perturbed Einstein tensor, viz. $\delta G^{\mu}_{\nu}$, 
corresponding to the line element~(\ref{eq:flesp}) above can be 
evaluated to be~\cite{bassett-2006}
\bseq
\br
\delta G_{0}^{0} 
&=& -6\, H\, \l({\dot \Psi}+H\, \Phi\r) 
+ \l(\f{2}{a^{2}}\r)\, \n^2 \Psi,\\
\delta G^{0}_{i} 
&=& 2\; \n_{i}\l({\dot \Psi}+H\, \Phi\r),\\
\delta G^{i}_{j} 
&=& -2\; \biggl[{\ddot \Psi}+ H\, \l(3\, {\dot \Psi}+{\dot \Phi}\r)\nn\\
& &\qquad\quad
+\l(2\, {\dot H}+3\, H^2\r)\, \Phi
+\l(\f{1}{a^{2}}\r)\, \n^{2}{\cal D}\biggr]\;\delta^{i}_{j}\nn\\
& &\qquad\quad
+ \l(\f{1}{a^{2}}\r)\,\n^{i}\, \n_{j}{\cal D}.
\er 
\eseq
where ${\cal D}=\l(\Phi-\Psi\r)$.


\subsubsection{Equations of motion}

The only sources of perturbations that I shall consider in this review 
will be scalar fields and perfect fluids.
As I shall illustrate in the next section, scalar fields do not possess 
any anisotropic stress at the linear order in the perturbations.
I shall assume that the perfect fluids that I consider do not contain 
any anisotropic stresses either. 
Under these conditions, the perturbed stress-energy tensor associated 
with such sources can be expressed as follows: 
\beq
\delta T^{0}_{0}=\delta \rho,\quad
\delta T^{0}_{i}=\l(\n_{i}\,\delta \sigma\r)\quad{\rm and}\quad
\delta T^{i}_{j}=-\delta p\; \delta^{i}_{j},
\label{eq:pstsp}
\eeq
where the quantities $\delta\rho$, $\delta \sigma$, and $\delta p$ 
are the scalar quantities that denote the perturbations in the energy 
density, the momentum flux, and the pressure, respectively.
The first order Einstein's equations, viz. $\delta G^{\mu}_{\nu}
=\l(8\,\pi\, G\r)\, \delta T^{\mu}_{\nu}$, then lead to the required
set of equations governing the perturbations.
It is clear from the form of $\delta G^{i}_{j}$ above that, in the
absence of anisotropic stresses, the corresponding Einstein equation
leads to: $\Phi=\Psi$.
In such a case, the remaining three Einstein equations simplify 
to~\cite{bassett-2006}
\bseq
\label{eq:foeesp}
\br
\!\!\!\!\!\!\!\!\!\!\!\!
-3\, H\, \l({\dot \Phi}+H\, \Phi\r) 
+ \l(\!\f{1}{a^{2}}\!\r) \n^2 \Phi\!\!
&=&\!\! \l(4\,\pi\,G\r)\, \delta \rho,\\
\!\!\!\!\!\!\!\!\!\!\!\!
\n_{i}\l({\dot \Phi}+H\, \Phi\r)\!\!
&=&\!\!\l(4\,\pi\,G\r)\, \l(\n_{i}\,
\delta \sigma\r),\;\;\\
\!\!\!\!\!\!\!\!\!\!\!\!
{\ddot \Phi}+ 4\,H\, {\dot \Phi}
+\l(2\, {\dot H}+3\, H^2\r)\, \Phi\!\!
&=&\!\!\l(4\,\pi\,G\r)\, \delta p.
\er
\eseq
The first and the third of these Einstein equations can be combined 
to lead to the following differential equation for the Bardeen 
potential~$\Phi$~\cite{mukhanov-1992,bardeen-1980}:
\br
& &\!\!\!\!\!\!\!\!\!\!\!\!\!\!\!\!
\Phi^{\prime\prime}
+3\, {\mathcal H}\, \l(1+c_{_{\rm A}}^2\r)\, \Phi^{\prime}
-c_{_{\rm A}}^2\, \nabla^{2}\Phi\nn\\
& &\!\!\!\!\!\!\!\!\,
+ \l[2\, {\mathcal H}'+ \l(1+3\, c_{_{\rm A}}^2\r)\,
{\mathcal H}^2\r]\, \Phi
= \l(4\,\pi\, G\, a^2\r)\, \delta p^{_{\rm NA}},\;\;
\label{eq:emPhi}
\er
where ${\cal H}=(a'/a)$ is the conformal Hubble parameter.
In arriving at the above equation, I have changed over to 
the conformal time coordinate, and have made use of the 
standard relation~\cite{gordon-2001}
\beq
\delta p= \l(c_{_{\rm A}}^2\, \delta\rho 
+  \delta p^{_{\rm NA}}\r),
\label{eq:deltapgen}
\eeq
where $c_{_{\rm A}}^2\equiv\l(p'/\rho'\r)$ denotes the adiabatic 
speed of the perturbations, and $\delta p^{_{\rm NA}}$ represents 
the non-adiabatic pressure perturbation.


\subsubsection{A conserved quantity at super-Hubble 
scales}\label{subsec:cR}

Consider the following quantity combination of the Bardeen 
potential~$\Phi$ and its time 
derivative~\cite{durrer-2008,mukhanov-1992,lukash-1980,lyth-1985}:
\beq
{\cal R}=\Phi + \l(\frac{2\, \rho}{3\, {\cal H}}\r)\,
\l(\frac{\Phi'+{\cal H}\, \Phi}{\rho+p}\r),\label{eq:R}
\eeq
a quantity that is referred to as the curvature 
perturbation\footnote{It is called so since it is 
proportional to the local three curvature on the 
spatial hypersurface~\cite{riotto-2002,giovannini-2005}.}. 
Upon substituting this expression in Eq.~(\ref{eq:emPhi}) that 
describes the evolution of the potential~$\Phi$, and making use 
of the background equations~(\ref{eq:fe}), one obtains that, in
Fourier space, 
\beq
{\cal R}_{k}'=\l(\frac{{\cal H}}{{\cal H}^{2}-{\cal H}'}\r)\;
\l[\l(4\,\pi\, G\, a^{2}\r)\, \delta p^{_{\rm NA}}_{k}
-c_{_{\rm A}}^{2}\; k^{2}\,\Phi_{k}\r].
\label{eq:emR'}
\eeq
(Henceforth, the sub-scripts~$k$ shall refer to the wavenumber of 
the Fourier modes of the perturbations.)
Note that, at super-Hubble scales, wherein the physical wavelengths 
of the perturbations are much larger than the Hubble radius [i.e.
when $(k/a\, H)=(k/{\cal H}) \ll 1$], the term $\l(c_{_{\rm A}}^2\, 
k^{2}\, \Phi_{k}\r)$ can be neglected. 
If one further assumes that no non-adiabatic pressure perturbations 
are present (i.e. $\delta p^{_{\rm NA}}=0$), say, as in the case of 
ideal fluids, then the above equation implies that 
${\cal R}_{k}^{\prime} \simeq 0$ at super-Hubble scales.
In other words, when the perturbations are adiabatic, the curvature 
perturbation ${\cal R}_{k}$ in conserved when the modes are well 
outside the Hubble radius.


\subsubsection{Evolution of the Bardeen potential at super-Hubble scales}

Let us now make use of the conservation of curvature perturbation to 
understand how the Bardeen potential evolves at super-Hubble scales
during the radiation and matter dominated eras.
If I now define~\cite{mukhanov-2005,mukhanov-1992,martin-2004}
\beq
\Phi =\l({\cal H}/a^2\, \theta\r)\, {\cal U},
\eeq
where
\beq
\theta
=\l[\f{{\cal H}^2}{\l({\cal H}^{2}-{\cal H}^{\prime}\r)\, 
a^{2}}\r]^{1/2},\label{eq:theta}
\eeq
then, I find that, in Fourier space, the equation~(\ref{eq:emPhi}) that 
governs the Bardeen potential reduces to
\beq
{\cal U}_{k}^{\prime\prime}+\l[c_{_{\rm A}}^2\, k^{2}
- \l(\f{\theta^{\prime\prime}}{\theta}\r)\r]\, {\cal U}_{k}
= \l(\f{4\,\pi\, G\, a^{4}\, \theta}{\cal H}\r)\, 
\delta p^{_{\rm NA}}_{k}.\label{eq:Uk} 
\eeq

In the absence of non-adiabatic pressure perturbations (i.e. when 
$\delta p^{_{\rm NA}}=0$), the above differential equation for 
${\cal U}_{k}$ simplifies to
\beq
{\cal U}_{k}^{\prime\prime}+ \l[c_{_{\rm A}}^2\, k^{2}
- \l(\frac{\theta^{\prime\prime}}{\theta}\r)\r]\, {\cal U}_{k} 
= 0.
\eeq
And, in the super-Hubble limit, say, as $k\to 0$, the 
general solution to this differential equation can be 
written as~\cite{mukhanov-2005,martin-2004}
\beq
{\cal U}_{k}(\eta)\simeq C_{_{\rm G}}(k)\; \theta(\eta)
\int\limits^{\eta}\f{\d{{\tilde \eta}}}{\theta^2({\tilde \eta})}
+C_{_{\rm D}}(k)\; \theta(\eta),
\eeq
where the coefficients $C_{_{\rm G}}$ and $C_{_{\rm D}}$ are 
$k$-dependent constants that are determined by the initial 
conditions imposed at early times.
The corresponding Bardeen potential~$\Phi_{k}$ is then given by
\beq
\Phi_{k}(\eta)\simeq C_{_{\rm G}}(k)\; \l(\f{\cal H}{a^{2}(\eta)}\r)\,
\int\limits^{\eta}\f{\d{{\tilde \eta}}}{\theta^2({\tilde \eta})}
+C_{_{\rm D}}(k)\; \l(\f{\cal H}{a^{2}(\eta)}\r).
\label{eq:Phikshs}
\eeq

Consider the power law expansion~(\ref{eq:ple}) that we had
discussed earlier.
Such an expansion can be expressed in terms of the conformal time 
coordinate as follows:
\beq
a(\eta)=\l(-{\bar {\cal H}}\, \eta \r)^{(\gamma+1)},
\label{eq:pleeta}
\eeq
where $\gamma$ and ${\bar {\cal H}}$ are constants given by
\beq
\gamma = -\l(\f{2\,q-1}{q - 1}\r)
\quad{\rm and}\quad
{\bar {\cal H}} = \l[(q - 1) \, a_{1}^{1/q}\r].
\label{eq:gammaq}
\eeq
In addition to power law inflation (which corresponds to $q>1$), 
the above scale factor describes the radiation and the matter 
dominates epochs (corresponding to $q=(1/2)$ and $(2/3)$, 
respectively) as well.
During inflation, $\gamma \le -2$, with $\gamma=-2$ corresponding 
to exponential inflation (i.e. $q\to\infty$).
While $\gamma=0$ in the case of the radiation dominated epoch, 
$\gamma=1$ during matter domination.
Moreover, note that, the quantity ${\bar {\cal H}}$ is positive and 
$-\infty<\eta<0$ during inflation, whereas, during radiation 
and matter domination, ${\bar {\cal H}}$ is negative and $0<\eta<\infty$.
It is helpful to notice that, in all these instances, $\eta \to 0$ 
corresponds to the super-Hubble limit. 

In the background~(\ref{eq:pleeta}), the quantity $\theta$ that is
defined in Eq.~(\ref{eq:theta}) is found to be
\beq
\theta(\eta)=\l(\f{\gamma+1}{\gamma+2}\r)^{1/2}\,
\l(\f{1}{a(\eta)}\r), 
\eeq
so that, at super-Hubble scales, one has [cf. Eq.~(\ref{eq:Phikshs})]
\br
\Phi_{k}(\eta)
&\simeq& C_{_{\rm G}}(k)\, \l[\f{3\, (w+1)}{3\, w+5}\r]\nn\\
& &\qquad\quad
+\; C_{_{\rm D}}(k)\, 
\l[\f{2/(3\, w+1)}{{\bar {\cal H}}^{[2\, (\gamma+1)]}\;
\eta^{(2\, \gamma+3)}}\r],\qquad
\er
where $w$ is the following equation of state parameter:
\beq
w\equiv(p/\rho)=\l[(1-\gamma)/3\,(1+\gamma)\r]
\eeq
that is a constant in power law expansion.
The first term in the above expression for $\Phi_{k}$ denotes the 
growing mode (which is actually a constant), while the second term
represents the decaying mode.
(Hence, the choice of sub-scripts ${\rm G}$ and ${\rm D}$ to the 
coefficient $C$.)
Demanding finiteness at very early times implies that the decaying 
mode has to be neglected, so that, at super-Hubble scales, I 
have~\cite{martin-2004}
\beq
\Phi_{k}(\eta)
\simeq C_{_{\rm G}}(k)\, \l[\frac{3\,(w+1)}{3\, w+5}\r].
\label{eq:Phikgm}
\eeq
This quantity vanishes when $w=-1$, which corresponds to exponential 
expansion driven by the cosmological constant.
Therefore, it is often said that the cosmological constant does not
induce any metric perturbations.

Since $\Phi_{k}$ is a constant at super-Hubble scales, in this limit,
the curvature perturbation ${\cal R}_{k}$ [cf.~Eq.~(\ref{eq:R})] is 
given by
\beq
{\cal R}_{k} 
\simeq \l[\frac{3\, w+5}{3\, (w+1)}\r]\, \Phi_{k}
\simeq C_{_{\rm G}}(k),\label{eq:Rkshs}
\eeq
where I have made use of the expression~(\ref{eq:Phikgm}) for 
$\Phi_{k}$.
As ${\cal R}_{k}$ is conserved and $\Phi_{k}$ is a constant 
at super-Hubble scales in power law expansion, when the modes 
enter the Hubble radius during the radiation or the matter 
dominated epochs, the Bardeen potentials at entry are given 
by~\cite{riotto-2002}
\bseq
\br
\Phi_{k}^{_{\rm R}}
&\simeq & \l[\frac{3\,(w_{_{\rm R}}+1)}{3\, w_{_{\rm R}}+5}\r]\, 
{\cal R}_{k}
=\l(\frac{2}{3}\r)\, C_{_{\rm G}}(k),\\
\Phi_{k}^{_{\rm M}}
&\simeq &\l[\frac{3\, (w_{_{\rm M}}+1)}{3\, w_{{\rm M}}+5}\r]\, 
{\cal R}_{k}
=\l(\frac{3}{5}\r)\, C_{_{\rm G}}(k),\qquad
\er
\eseq
where I have made use of the fact that $w_{_{\rm R}}=(1/3)$ and 
$w_{_{\rm M}}=0$.
These expressions also imply that, at super-Hubble scales, $\Phi_{k}$ 
changes by a factor of $(9/10)$ during the transition from the radiation 
to the matter dominated epoch~\cite{dodelson-2003,riotto-2002}.

It is essentially the spectrum of the Bardeen potential when the modes 
enter the Hubble radius during the radiation and the matter dominated
epochs that determines the pattern of the anisotropies in the CMB and 
the structure of the universe at the largest scales that we observe 
today~\cite{dodelson-2003,weinberg-2008,durrer-2008}.
The quantity $C_{_{\rm G}}(k)$ that determines such a spectrum has 
to be arrived at by solving the equation~(\ref{eq:Uk}) exactly with
suitable initial conditions imposed in the early universe.
In the inflationary scenario, physically well motivated, quantum, initial 
conditions are imposed on the modes when they are deep inside the Hubble 
radius.
As I shall illustrate in the following section, under these conditions, it 
is the background dynamics during the inflationary regime that influences
the functional form of $C_{_{\rm G}}(k)$.


\subsection{Vector perturbations}

As I had done in the case of the scalar perturbations, I shall work 
in a particular gauge to describe the vector perturbations as well.
I shall choose a gauge wherein the Friedmann metric, when the 
vector perturbations have been included, is described by the 
line element~\cite{mukhanov-2005,weinberg-2008,durrer-2008}
\beq
\d s^{2}=\d t^{2}-a^{2}(t)\, \l[\delta_{ij}+\l(\n_{i}\,F_{j}
+\n_{j}\,F_{i}\r)\r]\, \d x^{i}\, \d x^{j}.
\eeq
(This corresponds to setting ${\cal S}_{i}$ to zero and choosing
${\cal F}_{i}\propto F_{i}$.)
In such a gauge, upon using the condition that $F_{i}$ is a divergence
free vector, the different components of the perturbed Einstein tensor 
are found to be
\bseq
\br
\!\!
\delta G^{0}_{0}\!&=&\! 0,\quad
\delta G^{0}_{i}=\l(\f{1}{2}\r)\, \l(\n^{2}\, {\dot F}_{i}\r),\\
\!\!
\delta G^{i}_{j}\! &=&\! -\l(\f{1}{2}\r)
\biggl[3\, H\, \l(\n_{i}\, {\dot F}_{j}+\n_{j}\, {\dot F}_{i}\r)\nn\\
& &\qquad\qquad\qquad\qquad\quad
+\l(\n_{i}\, {\ddot F}_{j}+\n_{j}\, {\ddot F}_{i}\r)\biggr].\qquad\;
\er
\eseq
In the absence of any vector sources, according to the first order 
Einstein's equations, the non-zero components $\delta G^{0}_{i}$ and 
$\delta G^{i}_{j}$ have to be equated to zero. 
These then immediately imply that the metric perturbation $F_{i}$ vanishes
identically.
In other words, no vector perturbations are generated in the absence of 
sources with vorticity~\cite{mukhanov-2005,bassett-2006}.


\subsection{Tensor perturbations}

Let us now turn to the case of the tensor perturbations.
Upon the inclusion of these perturbations, the Friedmann 
metric can be described by the line element~\cite{dodelson-2003,mukhanov-2005,weinberg-2008,durrer-2008}
\beq
\d s^{2}=\d t^{2}-a^{2}(t)\, \l(\delta_{ij}+ h_{ij}\r)\, 
\d x^{i}\, \d x^{j},
\eeq
where $h_{ij}$ is a symmetric, transverse and traceless tensor.
(Note that $h_{ij}\propto {\cal H}_{ij}$.)
As I had discussed, the transverse and traceless conditions reduce the
number of independent degrees of freedom of $h_{ij}$ to two.
These two degrees correspond to the two types of polarization associated 
with the gravitational waves.
It can be shown that, on imposing the transverse and 
the traceless conditions, the components of the perturbed
Einstein tensor corresponding to the above line element 
simplify to~\cite{mukhanov-2005}
\bseq
\br
\!\!
\delta G^{0}_{0}&=&\delta G^{0}_{i}=0,\\
\!\!
\delta G^{i}_{j}\!&=&\!-\l(\f{1}{2}\r)\,
\l({\ddot h}_{ij}+3\, H\, {\dot h}_{ij}
-\l(\f{1}{a^{2}}\r)\, \n^{2} h_{ij}\r).\qquad
\er
\eseq
In the absence of anisotropic stresses, one then arrives at the 
following differential equation describing the amplitude~$h$ of 
the gravitational 
waves~\cite{dodelson-2003,mukhanov-2005,weinberg-2008,durrer-2008}:
\beq
h''+2\, {\cal H}\, h' - \n^{2}\, h=0,
\label{eq:emh}
\eeq
where, for later use, I have expressed the equation in terms of
the conformal time coordinate.


\section{Generation of perturbations during inflation}\label{sec:gpdi}

As I have repeatedly pointed out, the striking feature of inflation is 
the fact that it provides a natural mechanism to generate the 
perturbations~\cite{mukhanov-1981,hawking-1982,starobinsky-1982,guth-1982}.
It is the quantum fluctuations associated with the inflaton that act as 
the primordial seeds for the inhomogeneities.
I had earlier alluded to the fact that it is the spectrum of the Bardeen 
potential that determines the pattern of the anisotropies in the 
CMB and the formation of structures.
Since the curvature perturbation is proportional to the Bardeen 
potential at super-Hubble scales [cf.~Eq.~(\ref{eq:Rkshs})], the
primary quantity of interest is the spectrum of curvature 
perturbations generated during inflation.
The inflaton being a scalar source, does not degenerate any vector
perturbations~\cite{weinberg-2008,durrer-2008}.
However, as I have discussed, gravitational waves are generated even in 
the absence of sources~\cite{grishchuk-1974,starobinsky-1979,rubakov-1982}.
The primordial gravitational waves are also important 
because they too leave their own distinct imprints on 
the CMB~\cite{liddle-1999,dodelson-2003,durrer-2008}.
In this section, I shall first obtain the equation of motion governing the
curvature perturbation when the universe is dominated by the inflaton.
I shall then quantize the curvature and the tensor perturbations, impose 
vacuum initial conditions, and evaluate the scalar and the tensor spectra 
in super-Hubble limit for the cases of power law and slow roll inflation. 


\subsection{Equation of motion for the curvature perturbation}

As before, let $\phi$ denote the homogeneous scalar field.
Also, let $\delta \phi$ denote the perturbation in the field. 
It is then straightforward to show that, in the metric~(\ref{eq:flesp}), 
the components of the perturbed stress-energy tensor 
[cf.~Eqs.~(\ref{eq:setsf}) and (\ref{eq:pstsp})] associated with the 
scalar field can be expressed as
\bseq
\br
\delta T^{0}_{0}
&=& \l({\dot \phi}\; \dot{\delta \phi} 
- {\dot \phi}^{2}\; \Phi 
+ V_{\phi}\, \delta \phi\r)=\delta \rho\\
\delta T^{0}_{i}
&=& \n_{i}\,\l({\dot \phi}\; \delta\phi\r)
=\n_{i}\l(\delta \sigma\r),\\
\delta T^{i}_{j}
&=& -\l({\dot \phi}\; \dot{ \delta \phi} 
- {\dot \phi}^{2}\; \Phi 
- V_{\phi}\, \delta \phi\r)\; \delta^{i}_{j}
=- \delta p\; \delta^{i}_{j}.\qquad\quad
\er
\eseq
Evidently, the scalar field does not possess any anisotropic stress.
As a result, $\Phi=\Psi$ during inflation.
On substituting the above expressions for $\delta\rho$, $\delta \sigma$, 
and $\delta p$ in the first order Einstein equations~(\ref{eq:foeesp}) 
governing the scalar perturbations, one can arrive at following equation 
for the Bardeen potential:
\br
& &\!\!\!\!\!\!\!\!\!\!\!\!\!\!\!\!
\Phi^{\prime\prime}
+3\, {\mathcal H}\, \l(1+c_{_{\rm A}}^2\r)\, \Phi^{\prime}
-c_{_{\rm A}}^2\, \nabla^{2}\Phi\nn\\
& &\!\!\!\!\!\!\!\!\,
+ \l[2\, {\mathcal H}'+ \l(1+3\, c_{_{\rm A}}^2\r)\,
{\mathcal H}^2\r]\, \Phi
= \l(1-c_{_{\rm A}}^2\r)\,\nabla^{2}\Phi.\;\;
\label{eq:emPhisf}
\er

Upon comparing this equation for $\Phi$ with the general 
equation~(\ref{eq:emPhi}), it is evident that the non-adiabatic 
pressure perturbation associated with the inflaton is given by
\beq
\delta p^{_{\rm NA}}=\l(\f{1-c_{_{\rm A}}^2}{4\,\pi\, G\, a^2}\r)\,
\nabla^{2}\Phi.\label{eq:deltapNAsf}
\eeq
In such a case, the equation~(\ref{eq:emR'}) that describes the 
evolution of the curvature perturbation simplifies to
\beq
{\cal R}_{k}'
=  -\l(\f{\cal H}{{\cal H}^{2}-{\cal H}'}\r)\, 
\l(k^{2}\; \Phi_{k}\r).
\eeq
On differentiating this equation again with respect to time 
and, on using the background equations~(\ref{eq:HHdot}), the 
definition~(\ref{eq:R}) and the Bardeen equation~(\ref{eq:emPhisf}), 
one obtains the following equation of motion governing the Fourier 
modes of the curvature perturbation induced by the scalar
field~\cite{durrer-2008,bassett-2006}:
\beq
{\cal R}_{k}''+2\, \l(\frac{z'}{z}\r)\, {\cal R}_{k}'
+k^{2}\, {\cal R}_k=0,\label{eq:emRk}
\eeq
where the quantity $z$ is given by
\beq
z = \l(a\, {\dot \phi}/H\r)=\l(a\, \phi'/{\cal H}\r).
\label{eq:z}
\eeq

It is useful to introduce the Mukhanov-Sasaki variable~$v$ that is 
defined as~\cite{mukhanov-1985,sasaki-1986}
\beq
v=({\cal R}\, z).
\eeq
The Fourier modes of this variable, say, $v_{k}$, satisfy the 
differential equation
\beq
v_{k}^{\prime\prime} 
+  \l[k^2 - \l(\frac{z''}{z}\r)\r] v_{k} = 0.
\label{eq:msek}
\eeq


\subsection{Quantization of the perturbations and the definition
of the power spectra}

On quantization, the homogeneity of the Friedmann background
allows us to express the curvature perturbation ${\cal R}$ in 
terms of the Fourier modes ${\cal R}_{k}$ [satisfying 
Eq.~(\ref{eq:emRk})] as follows~\cite{martin-2004}:
\beq 
{\hat {\cal R}}\l(\eta, {\bf x}\r)\!
=\!\!\int\! \f{\d^{3}{\bf k}}{(2\,\pi)^{3/2}}
\l[{\hat a}_{\bf k}\, {\cal R}_{k}(\eta)\, e^{i\,{\bf k}\cdot{\bf x}}
+ {\hat a}_{\bf k}^{\dag}\, {\cal R}_{k}^{*}(\eta)\,
e^{-i\,{\bf k}\cdot{\bf x}}\r],
\label{eq:vdcmpstn}  
\eeq 
where the creation and the annihilation operators ${\hat a}_{\bf k}$
and ${\hat a}_{\bf k}^{\dag}$ obey the standard commutation relations. 
At the linear order in the perturbation theory that I am working 
in, the power spectrum as well as the statistical properties of 
the scalar perturbations are entirely characterized by the two 
point function of the quantum field~${\hat {\cal R}}$. 
(This is why the perturbations generated by inflation are often termed 
as Gaussian. 
However, basically, this happens to be true since we are restricting 
ourselves to the linear order in perturbation theory.
I shall briefly comment on possible deviations from Gaussianity in the
final section.)
The power spectrum of the scalar perturbations, say, 
${\cal P}_{_{\rm S}}(k)$, is given by the relation
\br
& &\!\!\!\!\!\!\!\!\!\!
\int\limits_{0}^{\infty}\frac{dk}{k}\, 
{\cal P}_{_{\rm S}}(k)
\equiv
\int \f{\d^{3}({\bf x}-{\bf x}')}{(2\, \pi)^{3}}\;
\langle 0 \vert {\hat {\cal R}}(\eta, {\bf x})\;
{\hat {\cal R}}(\eta, {\bf x'})\vert 0\rangle\nn\\
& &\qquad\qquad\qquad\qquad\quad\;\;\;
\times\, \exp -i\, \l[{\bf k}\cdot({\bf x}-{\bf x}')\r],\qquad
\label{eq:spsdfntn}
\er
where $\vert 0\rangle$ is the vacuum state defined as $\hat a_{\bf k}
\vert 0\rangle=0$ $\forall\; {\bf k}$. 
Using the decomposition~(\ref{eq:vdcmpstn}), the scalar perturbation 
spectrum can then be obtained to be~\cite{bassett-2006}
\beq  
{\cal P}_{_{\rm S}}(k)
=\l(\f{k^3}{2\,\pi^2}\r)\, 
\vert {\cal R}_{k}\vert^2
=\l(\frac{k^3}{2\,\pi^2}\r)\, 
\l(\f{\vert v_{k}\vert}{z}\r)^2.
\eeq 
The expression on the right hand side is to be evaluated at super-Hubble
scales [i.e. when $(k/a\, H)=(k/{\cal H})\ll 1$] when the curvature 
perturbation approaches a constant value\footnote{Earlier, 
in Subsec.~\ref{subsec:cR}, I had illustrated that, if the non-adiabatic 
pressure perturbation $\delta p^{_{\rm NA}}$ can be neglected, the 
curvature perturbation is conserved at super-Hubble scales. 
It can be shown that the non-adiabatic pressure perturbation 
associated with the inflaton [cf. Eq.~(\ref{eq:deltapNAsf})] 
decays exponentially in the super-Hubble limit and, hence, can be 
ignored~\cite{leach-2001,rajeev-2007}.
As a matter of fact, it is for this reason that the scalar 
perturbations produced by single scalar fields are usually referred 
to as adiabatic.}.

The tensor perturbations can be quantized in a similar fashion as the
curvature perturbation, and the corresponding spectrum can also be
defined in the same way.
If we write $h=(u/a)$, then, in Fourier space, the equation~(\ref{eq:emh})
describing the tensor perturbations reduces to
\beq
u_{k}''+\l[k^{2}-\l(\f{a''}{a}\r)\r]\, u_{k}=0.\label{eq:uk}
\eeq
And, it is helpful to note that this equation is essentially the same
as the Mukhanov-Sasaki equation~(\ref{eq:msek}) with the quantity~$z$
replaced by the scale factor~$a$.
The tensor perturbation spectrum, say, ${\cal P}_{_{\rm T}}(k)$, can 
then be expressed in terms of the modes $h_{k}$ and $u_{k}$ as follows:
\beq  
{\cal P}_{_{\rm T}}(k)
=2\,\left(\frac{k^3}{2\, \pi^2}\right)\, \vert h_{k}\vert^2
=2\,\left(\frac{k^3}{2\, \pi^2}\right)\,
 \l(\f{\vert u_{k}\vert}{a}\r)^{2}
\eeq 
with the expressions on the right hand sides to be evaluated in the
super-Hubble limit, as in the scalar case.
The additional factor of two in the above tensor spectrum has been 
included to take into account the two states of polarization of the 
gravitational waves.

The scalar and the tensor spectral indices are defined 
as~\cite{bassett-2006}
\beq
n_{_{\rm S}}
=1+\l(\f{\d\, {\rm ln}\, {\cal P}_{_{\rm S}}}{\d\, {\rm ln}\, k}\r)
\quad{\rm and}\quad
n_{_{\rm T}}
=\l(\f{\d\, {\rm ln}\, {\cal P}_{_{\rm T}}}{\d\, {\rm ln}\, k}\r).
\label{eq:si}
\eeq
Notice the difference in the definition of these two quantities.
Conventionally, a scale invariant scalar spectrum corresponds to
$n_{_{\rm S}}=1$, while such a tensor spectrum is described by 
$n_{_{\rm T}}=0$.
Finally, the tensor-to-scalar ratio $r$ is defined as 
follows~\cite{bassett-2006}:
\beq
r(k) 
\equiv \l(\f{{\cal P}_{_{\rm T}}(k)}{{\cal P}_{_{\rm S}}(k)}\r).
\eeq
As I shall discuss below, the scalar spectral index $n_{_{\rm S}}$ 
and the tensor-to-scalar ratio $r$ happen to be important inflationary
parameters that can be constrained by the observations.


\subsection{The Bunch-Davies initial conditions}

As I have mentioned, during inflation, the initial conditions on the 
perturbations are imposed when the modes are well inside the Hubble
scale [i.e. when $(k/a\, H)=(k/{\cal H})\gg 1$].
It is clear from Eqs.~(\ref{eq:msek}) and~(\ref{eq:uk}) that, in such 
a sub-Hubble limit, the scalar and tensor modes $v_{k}$ and $u_{k}$ do 
not feel the curvature of the spacetime and, hence, the solutions to 
these modes behave in the following Minkowskian form: $e^{\pm (i\,k\,
\eta)}$.
The assumption that the scalar and tensor perturbations are in the vacuum 
state then requires that $v_{k}$ and $u_{k}$ are positive frequency modes 
at sub-Hubble scales, i.e. they have the asymptotic
form~\cite{liddle-1999,mukhanov-2005} 
\beq
\lim _{\l(k/{\cal H}\r)\to \infty} \l(v_{k}(\eta), u_{k}(\eta)\r)
\to \l(\frac{1}{\sqrt{2k}}\r)\, {\rm e}^{-ik\eta}.
\label{eq:bdic}
\eeq
It should be pointed out that the vacuum state associated with the modes
that exhibit such a behavior is often referred to in the literature as 
the Bunch-Davies vacuum~\cite{bunch-1978}.


\subsection{The scalar and the tensor power spectra in power
law and slow roll inflation}\label{sec:sts}

In this section, my main goal will be to arrive at the scalar and 
the tensor perturbation spectra in slow roll inflation.
However, deriving the perturbation spectra in the case of power 
law inflation proves to be highly instructive in understanding the 
computation of the slow roll spectra. 
Therefore, before I derive the spectra in slow roll inflation, I shall 
discuss the power law case.


\subsubsection{The perturbation spectra in power law inflation}

In power law inflation, from the expressions~(\ref{eq:ple}) 
and~(\ref{eq:phitple}), it is clear that
\beq
z=\l(\f{a\, {\dot \phi}}{H}\r)= \sq{(2/q)}\; \MP\, a.
\eeq
Upon using the scale factor~(\ref{eq:pleeta}), the solution to the 
Mukhanov-Sasaki equation~(\ref{eq:msek}) that satisfies the initial
condition~(\ref{eq:bdic}) is found to 
be~\cite{abbott-1984,lyth-1992,martin-1998,sriram-2005}
\beq
v_{k}(\eta)
= \l(\frac{-\pi\, \eta}{4}\r)^{1/2}\, 
{\rm e}^{i\,\l[\nu + (1/2)\r]\, (\pi/2)}\;\; 
H^{(1)}_{\nu}\l(-k\eta\r),\label{eq:spli}
\eeq
where $\nu=-\l[\gamma+(1/2)\r]$, and $H_{\nu}^{(1)}$ is the Hankel 
function of the first kind and of order $\nu$. 
Since, in such a power law case,
\beq
\l(\f{z''}{z}\r)=\l(\f{a''}{a}\r),
\eeq
evidently, the tensor mode $u_{k}$ will be described by the same 
solution as the one for $v_{k}$ above.
As a result, barring an overall constant factor, the scalar and the 
tensor spectra will be of the same form.

Upon using the series expansion of the Hankel function in the
solution~(\ref{eq:spli}) and the expression~(\ref{eq:pleeta}),
it can readily shown that the curvature perturbation ${\cal R}_{k}$ 
and the tensor amplitude $h_{k}$ approach a constant value in the 
super-Hubble limit [i.e. as $\l(-k\,\eta\r) \to 0$], as expected. 
The scalar and the tensor power spectra evaluated 
at super-Hubble scales can be written 
as~\cite{abbott-1984,lyth-1992,martin-1998,sriram-2005}
\beq
{\cal P}_{_{\rm S(T)}}(k) 
= {\sf A}_{_{\rm S(T)}}\; {\bar {\cal H}}^{2}\;
\l(\f{k}{{\bar {\cal H}}}\r)^{2\, (\gamma+2)}.
\label{eq:pspli}
\eeq
The quantities ${\sf A}_{_{\rm S}}$ and ${\sf A}_{_{\rm T}}$
are given by 
\bseq
\br
{\sf A}_{_{\rm S}}
&=&\l[\f{\gamma+1}{16\, \pi^{3}\, (\gamma+2)\, \MP^{2}}\r]\,
\l(\f{\vert\Gamma(\nu)\vert^{2}}{2^{(2\, \gamma+1)}}\r),\\
{\sf A}_{_{\rm T}}
&=&\l(\f{1}{\pi^{3}\, \MP^{2}}\r)\,
\l(\f{\vert\Gamma(\nu)\vert^{2}}{2^{(2\, \gamma+1)}}\r).
\er
\eseq
where $\Gamma(\nu)$ is the Gamma function and, as seems to be the 
convention~\cite{dodelson-2003,bassett-2006}, I have included by 
hand, a factor of $(4/\MP^{2})$ in ${\sf A}_{_{\rm T}}$.
Clearly, the spectral indices are constants, and are given by
\beq
\l(n_{_{\rm S}}-1\r)=n_{_{\rm T}}
=\l[2\, (\gamma+2)\r]=-\l(\f{2}{q-1}\r),\label{eq:nspli}
\eeq
where I have made use of the relation~(\ref{eq:gammaq}).
The resulting tensor-to-scalar ratio is also a constant, and is 
found to be
\beq
r=\l[\f{16\, (\gamma+2)}{\gamma+1}\r]
=\l(\frac{16}{q}\r).\label{eq:rpli}
\eeq
These results point to the fact that the scalar and tensor 
spectra turn more and more scale invariant as the scale factor
approaches exponential inflation (i.e. as $q\to \infty$).


\subsubsection{The power spectra in slow roll inflation}

Let us now evaluate the scalar and tensor spectra in slow roll 
inflation.
Using equation~(\ref{eq:Hdot}) and the expression~(\ref{eq:hsrp1}) 
for the first Hubble slow roll parameter $\epsilon_{_{\rm H}}$, the 
quantity $z$ defined in equation~(\ref{eq:z}) can be written as
\beq
z = \sqrt{2}\; \MP\, \l(a\, \sqrt{\epsilon_{_{\rm H}}}\r).
\eeq
Also, note that the equations~(\ref{eq:hsrp}) defining the two Hubble
slow roll parameters $\epsilon_{_{\rm H}}$ and $\delta_{_{\rm H}}$ can 
be expressed in terms of the conformal time coordinate as follows:
\beq
\epsilon_{_{\rm H}}
= 1 - \l(\f{{\cal H}'}{{\cal H}^2}\r)
\quad{\rm and}\quad
\delta_{_{\rm H}} 
= \epsilon_{_{\rm H}}  
- \l(\f{\epsilon_{_{\rm H}}'}{2\, {\cal H}\, \epsilon_{_{\rm H}}}\r).
\label{eq:hsreta}
\eeq
Using these expressions for $z$ and the HSR parameters, the term 
$(z''/z)$ that appears in the Mukhanov-Sasaki equation~(\ref{eq:msek}) 
can be written as~\cite{lidsey-1997,stewart-1993,hwang-1996}
\beq
\l(\!\frac{z''}{z}\!\r)
={\cal H}^2
\l[2 - \epsilon_{_{\rm H}} 
+ \l(\epsilon_{_{\rm H}} - \delta_{_{\rm H}}\r)\, 
\l(3 - \delta_{_{\rm H}}\r) 
+ \l(\!\f{\epsilon_{_{\rm H}}' - \delta_{_{\rm H}}'}{\cal H}\!\r)\r].
\label{eq:zppz}
\eeq
From the definition of $\epsilon_{_{\rm H}}$ above, it can also be
established that 
\beq
\l(\frac{a''}{a}\r)={\cal H}^2\, \l(2 - \epsilon_{_{\rm H}}\r).
\label{eq:appa}
\eeq

Let us now rewrite the expression~(\ref{eq:hsreta}) above for 
$\epsilon_{_{\rm H}}$ as follows: 
\beq
\eta = - \int \l(\frac{1}{1 - \epsilon_{_{\rm H}}}\r)\, 
\d\l(\frac{1}{\cal H}\r). 
\eeq
On integrating this expression by parts, and using the above definition 
of $\delta_{_{\rm H}}$, one obtains that
\beq
\eta 
= - \l[\f{1}{(1 - \epsilon_{_{\rm H}})\, {\cal H}}\r]\,
- \int \l[\f{2\, \epsilon_{_{\rm H}}\, 
\l(\epsilon_{_{\rm H}} - \delta_{_{\rm H}}\r)}
{(1 - \epsilon_{_{\rm H}})^3}\r]\, \d\l(\frac{1}{\cal H}\r). 
\eeq
At the leading order in the slow roll approximation [cf. 
Eq.~(\ref{eq:sra})], the second term can be ignored and, 
at the same order, one can assume $\epsilon_{_{\rm H}}$ to 
be a constant.
Therefore, we have
\beq
{\cal H} 
\simeq - \l[\f{1}{\l(1 - \epsilon_{_{\rm H}}\r)\, \eta}\r].
\label{eq:calHsr}
\eeq
If we now use this expression for ${\cal H}$ in the
expressions~(\ref{eq:zppz}) and~(\ref{eq:appa}), then, at the 
leading order in the slow roll approximation, one gets that
\bseq
\br
\l(\f{z''}{z}\r)
&\simeq& \l(\f{2+6\, \epsilon_{_{\rm H}} 
-3\, \delta_{_{\rm H}}}{\eta^{2}}\r),\\
\l(\f{a''}{a}\r)
&\simeq& \l(\f{2+3\,\epsilon_{_{\rm H}}}{\eta^{2}}\r),
\er
\eseq
with the slow roll parameters treated as constants.
It is then clear from Eqs.~(\ref{eq:msek}) and~(\ref{eq:uk})
that the solutions to the variables $v_{k}$ and $u_{k}$ will 
again be given in terms of Hankel functions as in the power 
law case~(\ref{eq:spli}), with the quantity $\nu$ now given 
by
\beq
\nu_{_{\rm S}} 
\simeq  \l[\l(\f{3}{2}\r) 
+2\, \epsilon_{_{\rm H}} -\delta_{_{\rm H}}\r]
\quad{\rm and}\quad
\nu_{_{\rm T}} 
\simeq  \l[\l(\f{3}{2}\r) + \epsilon_{_{\rm H}}\r],
\eeq
where the subscripts $S$ and $T$ refer to the scalar and tensor 
cases. 

It now remains to evaluate the two spectra in the super-Hubble limit.
In this limit [i.e. as $(-k\,\eta)\to 0$], upon expanding the Hankel 
function as a series about the origin, the scalar and the tensor 
spectra can be expressed as~\cite{bassett-2006,stewart-1993}
\bseq
\br
\!\!\!\!\!\!\!\!\!\!\!\!\!\!\!\!
{\cal P}_{_{\rm S}}(k)\!\!
&=&\!\! \l(\!\f{1}{32\, \pi^{2}\, \MP^{2}\;
\epsilon_{_{\rm H}}}\!\r)
\l[\f{\vert \Gamma(\nu_{_{\rm S}})\vert}{\Gamma(3/2)}\r]^{2}
\l(\!\f{k}{a}\!\r)^{2}\!
\l(\!\frac{-k\, \eta}{2}\!\r)^{(1-2\,\nu_{_{\rm S}})}\nn\\
&=&\!\! \l(\f{H^{2}}{2\, \pi\, {\dot \phi}}\r)^{2}\, 
\l[\f{\vert \Gamma(\nu_{_{\rm S}})\vert}{\Gamma(3/2)}\r]^{2}\nn\\
& &\qquad\qquad
\times\;\,2^{(2\,\nu_{_{\rm S}}-3)}\,
\l(1-\epsilon_{_{\rm H}}\r)^{(2\,\nu_{_{\rm S}}-1)},\\
{\cal P}_{_{\rm T}}(k)\!\!
&=&\!\! \l(\f{1}{2\, \pi^{2}\, \MP^{2}}\r)\, 
\l[\f{\vert \Gamma(\nu_{_{\rm T}})\vert}{\Gamma(3/2)}\r]^{2}
\l(\f{k}{a}\r)^{2}\, 
\l(\f{-k\, \eta}{2}\r)^{(1-2\,\nu_{_{\rm T}})}\nn\\
&=&\!\! \l(\f{2\,H^{2}}{\pi^{2}\,\MP^{2}}\r)\, 
\l[\f{\vert \Gamma(\nu_{_{\rm T}})\vert}{\Gamma(3/2)}\r]^{2}\nn\\
& &\qquad\qquad
\times\;\,2^{(2\,\nu_{_{\rm T}}-3)}\,
\l(1-\epsilon_{_{\rm H}}\r)^{(2\,\nu_{_{\rm T}}-1)},
\er
\eseq
where $H$ is the Hubble parameter, and the second equalities express 
the asymptotic values in terms of the values of the quantities at 
Hubble exit [i.e. at $(-k\,\eta)=(1-\epsilon_{_{\rm H}})^{-1}$].
(Also, I have multiplied the tensor spectrum by the factor of
$(4/\MP^2)$, as I had mentioned earlier.)
At the leading order in the slow roll approximation, the amplitudes 
of the scalar and the tensor spectra can easily be read off from the 
above expressions.
They are given by~\cite{bassett-2006}
\bseq
\label{eq:staisri}
\br
{\cal P}_{_{\rm S}}(k)
&\simeq& \l(\frac{H^2}{2\,\pi\, {\dot \phi}}\r)_{k=(a\,H)}^{2},\\
{\cal P}_{_{\rm T}}(k)
&\simeq& \l(\f{8}{\MP^{2}}\r)\,
\l(\frac{H}{2\, \pi}\r)_{k=(a\,H)}^{2},
\er
\eseq
with the sub-scripts on the right hand side indicating that the 
quantities have to be evaluated when the modes cross the Hubble 
radius.
Given a quantity, say, $y$, we can write~\cite{bassett-2006}
\br
\l(\f{\d y}{\d\, {\rm ln}\, k}\r)_{k=(a\, H)}
&=& \l(\f{\d y}{\d t}\r)\, \l(\f{\d t}{\d\, {\rm ln}\, a}\r)\,
\l(\f{\d\, {\rm ln}\, a}{\d\, {\rm ln}\, k}\r)_{k=(a\, H)}\nn\\
&=&\l(\f{\dot y}{H}\r)_{k=(a\, H)},\label{eq:he}
\er
where, in arriving at final expression, the following condition
has been used:
\beq
\l(\f{\d\, {\rm ln}\, a}{\d\, {\rm ln}\, k}\r)_{k=(a\,H)} 
\simeq 1,
\eeq 
as $H$ does not vary much during slow roll inflation.
Using the expressions~(\ref{eq:staisri}) for the power spectra, 
the definitions~(\ref{eq:si}) of the spectral indices, and the
relation~(\ref{eq:si}), one can easily show that
\beq
n_{_{\rm S}}
\simeq  \l(1-4\,\epsilon_{_{\rm H}}+2\,\delta_{_{\rm H}}\r)
\quad{\rm and}\quad
n_{_{\rm T}} 
\simeq -\l(2\, \epsilon_{_{\rm H}}\r).\label{eq:stsisri}
\eeq
These expressions unambiguously point to the fact the scalar and 
the tensor spectra that arise in slow roll inflation will be 
nearly scale invariant.
The tensor-to-scalar ratio in the slow roll limit is found to be
\beq
r \simeq \l(16\, \epsilon_{_{\rm H}}\r)
=-\l(8\, n_{_{\rm T}}\r)\label{eq:rsri}
\eeq
with the last equality often referred to as the consistency 
relation~\cite{lidsey-1997}.


\section{Comparison with the recent CMB observations}\label{sec:cwocim}

In this section, I shall briefly discuss as to how a power law
primordial spectrum compares with the recent observations of the 
anisotropies in the CMB.
I shall also indicate how the observations constrain some of the
models of inflation.

\begin{figure}[htbp!]
\begin{center}
\resizebox{240pt}{180pt}{\includegraphics{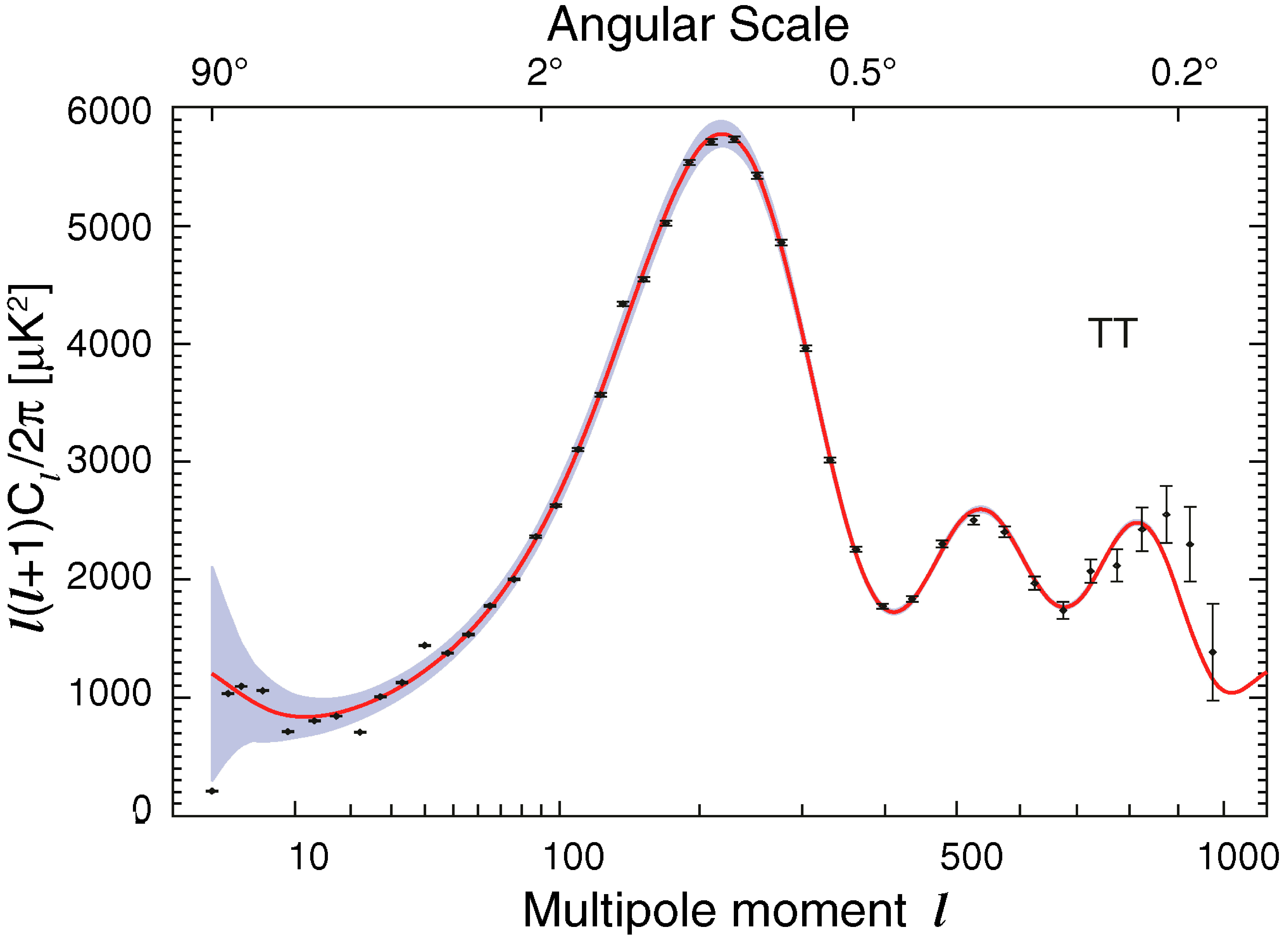}}
\caption{The angular power spectrum of the CMB temperature 
anisotropies from the WMAP $5$-year data (the black dots 
with error bars) and for the best fit concordant model, i.e. 
a spatially flat, $\Lambda$CDM model, with a power law 
primordial spectrum (the solid red curve).
The blue band denotes the statistical uncertainty, known as the 
cosmic variance~\cite{dodelson-2003,weinberg-2008,durrer-2008}.
The concordant model seems to fit the data rather well.
(This figure is from Ref.~\cite{hinshaw-2009}.)}
\label{fig:cl-wmap5}
\end{center}
\end{figure}

The nearly scale invariant scalar power spectrum and the rather small
tensor-to-scalar ratio that arise in slow roll inflation seems to 
be in good agreement with the observations of the anisotropies in 
the CMB.
In Fig.~\ref{fig:cl-wmap5}, the angular power spectrum of the CMB 
temperature anisotropies corresponding to the concordant cosmological 
model---viz. a spatially flat, $\Lambda$CDM model\footnote{The term 
$\Lambda$CDM model refers to the currently accepted composition of 
our universe, viz. about $72\%$ of dark energy, close to $23\%$ of 
cold (i.e. non-relativistic) dark matter, and roughly $5\%$ of baryons.
These numbers have been arrived at based on a 
variety of observations, including that of the 
CMB anisotropies~\cite{weinberg-2008,durrer-2008}.}, and a nearly 
scale invariant primordial spectrum---has been plotted as a function 
of the multipoles.
The figure also contains the data from the most recent observations of 
the CMB by the WMAP mission~\cite{hinshaw-2009}.
Visually, it is clear that the concordant model provides a reasonable 
fit to the data.
Detailed analysis of the data indicates that $n_{_{\rm S}}\simeq 0.96$, 
when the tensor contribution is completely ignored.
Whereas, it is found that $n_{_{\rm S}}\simeq 0.98$ when tensors are 
taken into account. 
The data also constrains the tensor-to-scalar ratio to $r<0.43$ at $95\%$ 
Confidence Level (CL)~\cite{komatsu-2009a}.

The slow roll approximation also enables specific inflationary 
models to be compared easily with the observations.
In order to do so, firstly, note that, in the slow roll limit, upon using 
the corresponding background equations~(\ref{eq:Hemsfsra}), the scalar and 
tensor spectral amplitudes~(\ref{eq:staisri}) can be expressed in terms of 
the potential $V(\phi)$ and its derivative as follows~\cite{bassett-2006}:
\bseq
\label{eq:stasriV}
\br
{\cal P}_{_{\rm S}}(k)
&\simeq& \l(\frac{1}{12\,\pi^2\, \MP^6}\r)\,
\l(\frac{V^3}{V_\phi^2}\r)_{k=(a\,H)},\\
{\cal P}_{_{\rm T}}(k)
&\simeq& \l(\frac{2}{3\, \pi^2}\r)\,
\left(\frac{V}{\MP^4}\r)_{k=(a\,H)}.
\er
\eseq
Secondly, during slow roll, the relations between the HSR and 
the PSR parameters can be obtained to be
\beq
\epsilon_{_{\rm H}}\simeq \epsilon_{_{\rm V}}
\quad{\rm and}\quad
\delta_{_{\rm H}}\simeq \l(\eta_{_{\rm V}}-\epsilon_{_{\rm V}}\r),
\label{eq:hsrpsr}
\eeq
and, as a result, for instance, $\epsilon_{_{\rm V}}\simeq 1$ 
indicates the end of inflation.
Let us now use these expressions to arrive at observational 
constraints on the parameters of a couple of inflationary models.

Observations indicate that the amplitude of the scalar 
perturbation associated with a mode that crossed the 
Hubble radius about $60$ $e$-folds before the end of 
inflation is about $\l(2\times10^{-9}\r)$, a constraint 
that is referred to as the COBE normalization~\cite{bunn-1996}.
Given a potential, the expressions~(\ref{eq:stsisri})--(\ref{eq:hsrpsr}) 
allow us to construct the scalar amplitude and spectral index
as well as the tensor-to-scalar ratio in terms of the inflaton.
Let me now focus on the large field models~(\ref{eq:lfm}) 
that I had discussed earlier. 
In these models, inflation ends (i.e. $\epsilon_{_{\rm V}}
\simeq 1$), when $\phi_{\rm end}\simeq [(n/\sqrt{2})\, \MP]$.
The value of the field at $N$ $e$-folds {\it before the end 
inflation}\/ can be obtained using the solution~(\ref{eq:slfm}), 
and is given by $\phi_{_{N}}\simeq (\sqrt{[(4\, N+n)\, n/2]}\; 
\MP)$.
Using these expressions, it is straightforward to show that, for $V_{0}
=(m^2/2)$ and $n=2$, the COBE normalization condition leads to the
constraint that $m\simeq (10^{-5}\, \MP$).
Similarly, for $V_{0}=(\lambda/4)$ and $n=4$, the condition leads to 
$\lambda\simeq 10^{-13}$.

\begin{figure}[htbp!]
\begin{center}
\resizebox{180pt}{300pt}{\includegraphics{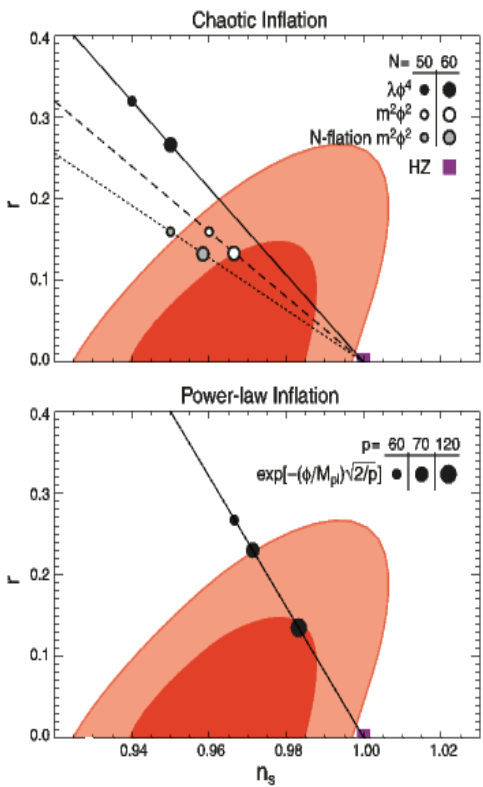}}
\caption{Joint constraints on the $n_{_{\rm S}}$-$r$~plane from 
the WMAP $5$ year data, and a couple of other data sets. 
The contours indicate the $68\%$ and $95\%$ CLs derived from the data.
The top panel clearly points to the fact that the data excludes the 
$n=4$ large field model at more than $95\%$~CL for $N<60$ (the solid 
line).
In contrast, the $n=2$ model (the dashed line) performs better, 
and falls at the boundary of the $68\%$ region for $N=60$.
(I should again stress that the number of $e$-folds have been 
counted {\it backwards from the end of inflation.})\/
The bottom panel indicates the behavior of power law inflation
in the slow roll limit (i.e. when $q\ll 1$).
It is evident that, while models with $q<60$ are outside of 
the $95\%$ region, models with $q\simeq 120$ lie close to 
the boundary of the $68\%$ region.
(Note that $p$ in the bottom panel is what I have called as~$q$. 
This figure is from Ref.~\cite{komatsu-2009a}, and I would refer 
the reader to this reference for any further details.)}
\label{fig:jcip}
\end{center}
\end{figure}

The CMB observations also lead to useful constraints on the 
models in the $n_{_{\rm S}}$-$r$~plane.
The various quantities that we have obtained above allow 
us to express the scalar spectral index and the tensor-to-scalar 
ratio in terms of the number of $e$-folds~$N$ {\it counted from 
the end inflation}.\/
For the large field models, the solution~(\ref{eq:slfm}) enables 
us to arrive at the following expressions in the slow roll
limit~\cite{bassett-2006}
\beq
n_{_{\rm S}}\simeq 1-\l[\f{2\,(n+2)}{4\,N+n}\r]
\quad{\rm and}\quad
r \simeq \l(\f{16\, n}{4\,N+n}\r).
\eeq
I have already shown that, for the case of power law 
inflation~(\ref{eq:ple}) described by the potential~(\ref{eq:plp}), 
$n_{_{\rm S}}$ and $r$ are constants given by Eqs.~(\ref{eq:nspli})
and (\ref{eq:rpli}), respectively.
These results for the power law case were {\it exact}.\/ 
In the slow roll limit (i.e. when $q\gg 1$), they reduce 
to~\cite{komatsu-2009a}
\beq
n_{_{\rm S}}\simeq 1-\l(\frac{2}{q}\r)
\quad{\rm and}\quad
r=\l(\f{16}{q}\r).
\eeq
These relations indicate that these models will be described by 
straight lines in the $n_{_{\rm S}}$-$r$~plane.
In Fig.~\ref{fig:jcip}, the joint constraints on the 
$n_{_{\rm S}}$-$r$~plane from the recent WMAP data (and a couple 
of other datasets) have been displayed~\cite{komatsu-2009a}. 
The behavior of a few inflationary models have been included in the
figure as well.
It is evident from the figure that, amongst the large field models,
the $n=2$ case performs better than the $n=4$ case.
Also, in the case of the power law inflation, it is found that $q<60$ 
is excluded at more than $95\%$ CL.
 

\section{Status and prospects of inflation}\label{sec:sp}

I shall finally close by briefly commenting on the status and 
some prospects of the inflationary paradigm.


\subsection{Profusion of inflationary models}

As a broad concept, inflation can certainly be considered 
as a success.
However, a specific model of inflation that can be 
satisfactorily embedded in a high energy theory still 
eludes us.
A plethora of inflationary models exist, quite a few of 
which are phenomenological in nature\footnote{I would 
urge the reader to take a look at the figure mentioned 
in Ref.~\cite{loim}, which lists the various inflationary 
models that have been considered in literature. 
The list is rather long, but, apparently, it is incomplete! 
It tells the tale.}.
Despite the enormous amount of effort, including attempts at
considering models that are not described by the canonical 
action, it will be fair to say that we do not yet have a 
satisfactory model.
The principal hurdle facing the idea of inflation is to 
construct a model that is well motivated from the high energy 
perspective, and also fits the observational data well.


\subsection{Features in the primordial spectrum}

In this review, I had restricted myself to discussions on 
slow roll inflation, which leads to a featureless and nearly 
scale invariant primordial spectrum that seems to agree well
with the recent observations of the CMB anisotropies.
Though the agreement is rather good, there also exist a few 
points at the lower multipoles of the observed CMB angular power 
spectrum which lie outside the cosmic variance associated with 
the concordant model.
Given these observations, a handful of model independent 
approaches have been constructed to recover the primordial 
power spectrum.
At the smaller scales, all these approaches arrive at a spectrum 
that is nearly scale invariant.
However, many of the approaches seem to unambiguously point to 
a sharp drop in power (along with a few distinct features) at 
the scales corresponding to the Hubble scale today.
If future observations support the presence of such features in 
the primordial spectrum, then it poses an interesting challenge 
to build inflationary models that lead to the required spectrum.
Conversely, the necessities of features will considerably restrict 
the class of allowed models of inflation.
(For various efforts in this direction, see the references listed
in Refs.~\cite{pi,tsr-pi}.)


\subsection{Deviations from Gaussianity}

I had earlier mentioned that the scalar perturbations generated 
during inflation are Gaussian in nature.
I had also clarified that this was primarily due to the fact that
we had confined ourselves to linear perturbation theory. 
Deviations from Gaussianity can arise when takes into account the 
perturbations at the higher orders~\cite{bartolo-2004a,bartolo-2004b}. 
However, the extent of the non-Gaussianity depends on a variety of 
reasons (for a recent discussion on the issue and further references, 
see Ref.~\cite{komatsu-2009b}). 
Interestingly, recent re-analysis of the WMAP $5$-year data seem to 
indicate sufficiently large non-Gaussianities (see, for instance,
Ref.~\cite{smith-2009}).
If future observations confirm such a large level of non-Gaussianity, 
then, it can result in a substantial tightening in the constraints on 
the inflationary models.
For example, canonical scalar field models that lead to a 
nearly scale invariant primordial spectrum contain only a 
small amount of non-Gaussianity and, hence, may cease to 
be viable~\cite{maldacena-2003}.
However, it is known that primordial spectra with features can lead 
to reasonably large non-Gaussianities~\cite{chen-2007a,chen-2007b}. 
Therefore, if non-Gaussianity indeed turns out to be large, then, 
either one may have to reconcile with the fact that the primordial 
spectrum contains features, or one possibly have to seriously
consider models described by non-canonical actions, some of which 
are known to result in large Gaussianities (see, for instance, 
Refs.~\cite{langlois-2008a,langlois-2008b}).

Hopefully, one or both of these aspects will help us arrive at a 
satisfactory model of inflation.


\begin{acknowledgments}
This review is partly based on a topical course titled `Origin and 
evolution of perturbations during inflation and reheating' that I 
had given at the Inter University Centre for Astronomy and Astrophysics 
(IUCAA), Pune, India in February 2009.
I would like to thank Tarun Souradeep for the initial suggestion to give 
a set of lectures, Kandaswamy Subramanian for the invitation to give 
the topical course, and IUCAA for the hospitality. 
I also wish to take this opportunity to acknowledge collaborations 
and/or discussions with Raul Abramo, Pravabati Chingangbam, Sudipta 
Das, Jinn-Ouk Gong, Rajeev Jain, Jerome Martin, T.~Padmanabhan, 
Raghavan Rangarajan, T.~R.~Seshadri, Tarun Souradeep, 
S.~Shankaranarayanan and Kandaswamy Subramanian, on these topics at 
different stages.
I would like to thank Sudipta Das and Rajeev Jain for comments 
on the manuscript.
I should also acknowledge the use of one figure each, viz. 
Figs.~\ref{fig:cl-wmap5} and~\ref{fig:jcip}, from 
Refs.~\cite{hinshaw-2009} and~\cite{komatsu-2009a}, respectively.
\end{acknowledgments}


\end{document}